\documentclass[aip,jcp,reprint]{revtex4-1}%
\usepackage{amsmath,amsfonts,amssymb,amsthm}
\usepackage{placeins}
\usepackage{tabularx,multirow}
\usepackage[T1]{fontenc}
\usepackage{graphicx}
\usepackage[english]{babel}
\usepackage{xcolor}
\usepackage{natbib}
\usepackage{amsmath}
\usepackage{amsfonts}
\usepackage{amssymb}
\usepackage{longtable}
\usepackage{mathtools}
\providecommand{\U}[1]{\protect\rule{.1in}{.1in}}
\usepackage{rotating}
\graphicspath{{./Figures/}{D:/Papers/Konstantin/EIE_extended_space/Figures/}
{C:/Users/Jiri/Dropbox/Papers/Chemistry_papers/2016/EIE_stochastic_DE/Revision/Figures/}{./}
}
\makeatletter
\providecommand{\tabularnewline}{\\}
\newcommand{\pz}{\phantom{0}}

\@ifundefined{textcolor}{}
{
\definecolor{BLACK}{gray}{0}
\definecolor{WHITE}{gray}{1}
\definecolor{RED}{rgb}{1,0,0}
\definecolor{GREEN}{rgb}{0,1,0}
\definecolor{BLUE}{rgb}{0,0,1}
\definecolor{CYAN}{cmyk}{1,0,0,0}
\definecolor{MAGENTA}{cmyk}{0,1,0,0}
\definecolor{YELLOW}{cmyk}{0,0,1,0}
}
\makeatother
\begin{document}
\title{Accelerating equilibrium isotope effect calculations: II. Stochastic
implementation of direct estimators}
\author{Konstantin Karandashev}
\email{konstantin.karandashev@alumni.epfl.ch}
\author{Ji\v{r}\'{\i} Van\'{\i}\v{c}ek}
\email{jiri.vanicek@epfl.ch}
\affiliation{Laboratory of Theoretical Physical Chemistry, Institut des Sciences et
Ing\'{e}nierie Chimiques, Ecole Polytechnique F\'{e}d\'{e}rale de Lausanne
(EPFL), CH-1015, Lausanne, Switzerland}
\date{\today}

\begin{abstract}
Path integral calculations of equilibrium isotope effects and isotopic
fractionation are expensive due to the presence of path integral
discretization errors, statistical errors, and thermodynamic integration
errors. Whereas the discretization errors can be reduced by high-order
factorization of the path integral and statistical errors by using centroid
virial estimators, two recent papers proposed alternative ways to completely
remove the thermodynamic integration errors: Cheng and Ceriotti [J. Chem. Phys. \textbf{141}, 244112 (2015)] employed a variant of free-energy perturbation called \ \textquotedblleft direct estimators\textquotedblright,
while Karandashev and Van\'{\i}\v{c}ek [J. Chem. Phys. \textbf{143}, 194104 (2017)] combined the thermodynamic integration with a stochastic change of mass and piecewise-linear umbrella biasing potential. Here we combine the former approach with the stochastic change of mass in order to decrease its statistical errors when applied to larger isotope effects, and perform a thorough comparison of different methods by computing isotope effects first on a harmonic model, and then on methane and methanium, where we evaluate all isotope effects of the form $\mathrm{CH}_{\mathrm{4-x}}\mathrm{D}_{\mathrm{x}}/\mathrm{CH}_{4}$ and $\mathrm{CH}_{\mathrm{5-x}}\mathrm{D}^{+}_{\mathrm{x}}/\mathrm{CH}^{+}_{5}$, respectively. We discuss thoroughly the reasons for a surprising behavior of the original method of direct estimators, which performed well for a much larger range of isotope effects than what had been expected previously, as well as some implications of our work for the more general problem of free energy difference calculations.

\end{abstract}
\maketitle

\section{Introduction}

Equilibrium isotope effect and a closely related concept of isotope
fractionation\cite{Ceriotti_Markland:2013,Marsalek_Tuckerman:2014,Cheng_Ceriotti:2014}
belong among the most useful experimental tools for uncovering the influence
of nuclear quantum effects on molecular
properties.\cite{Janak_Parkin:2003,Wolfsberg_Rebelo:2010} The equilibrium (or
thermodynamic) isotope effect measures the effect of isotopic substitution on
the equilibrium constant of a chemical reaction and is defined as the ratio of
equilibrium constants,
\begin{equation}
\mathrm{EIE}:=K^{(B)}/K^{(A)},
\end{equation}
where $A$ and $B$ are two isotopologues of the reactant. The equilibrium
constant can, in general, be evaluated as the ratio of the product and
reactant partition functions ($K=Q_{\text{prod}}/Q_{\text{react}}$), and
therefore every equilibrium isotope effect can be computed as a product of
several \textquotedblleft elementary\textquotedblright\ isotope effects
(IEs),
\begin{equation}
\mathrm{IE}=Q^{(B)}/Q^{(A)}, \label{eq:IE_definition}%
\end{equation}
given by the ratio of partition functions corresponding to different
isotopologues. In the following, we will discuss different approaches for
evaluating these elementary partition function ratios and call them
\textquotedblleft isotope effects\textquotedblright\ for short.

The standard textbook approach for computing the isotope effect assumes
separability of rotations and vibrations, rigid rotor approximation for the
rotations and harmonic approximation for the
vibrations,\cite{Urey:1947,Wolfsberg_Rebelo:2010} but we will focus our
discussion on a more rigorous method that avoids all three approximations and
treats the problem exactly. The method uses the Feynman path integral
formalism\cite{Feynman_Hibbs:1965, Chandler_Wolynes:1981, Ceperley:1995} to
transform the quantum partition function to a classical partition function of
the \textquotedblleft ring polymer,\textquotedblright\ a larger, but classical
system in an extended configuration space; the isotope effect can then be
evaluated via the thermodynamic integration\cite{Kirkwood:1935} with respect
to
mass.\cite{Vanicek_Aoiz:2005,Vanicek_Miller:2007,Zimmermann_Vanicek:2009,Zimmermann_Vanicek:2010,Perez_Lilienfeld:2011}
The main drawback of this approach is that the \textquotedblleft mass
integral\textquotedblright\ is evaluated by discretizing the mass and, as a
result, introduces a certain integration error. Although several elegant
tricks reduce this integration error
significantly,\cite{Ceriotti_Markland:2013,Marsalek_Tuckerman:2014} it can
never be removed completely if the integral is evaluated deterministically.

This disadvantage of thermodynamic integration led to the introduction of
several strategies that avoid the integration error altogether; these can be
classified into two main groups: The first group avoids discretizing the mass
integral by allowing the mass to take a continuous range of values during the
simulation;\cite{Perez_Lilienfeld:2011,Karandashev_Vanicek:2017a} this
approach is an example of a more general $\lambda$-dynamics
method\cite{Liu_Berne:1993,Kong_Brooks:1996,Guo_Kong:1998} for calculating free energy differences. Not only do these techniques eliminate the integration error, but they also tend to show faster statistical convergence than does standard thermodynamic integration;\cite{Bitetti-Putzer_Karplus:2003,Karandashev_Vanicek:2017a} this property is similar to the improvement achieved by parallel tempering.\cite{Swendsen_Wang:1986,Hukushima_Nemoto:1996,Predescu_Ciobanu:2005} The second group is based on the free energy perturbation,\cite{Zwanzig:1954,Oostenbrink:2009,Perez_Lilienfeld:2011,Ceriotti_Markland:2013} which can be derived by rewriting the ratio~(\ref{eq:IE_definition}) using the Zwanzig formula; the approach was proposed in Ref.~\onlinecite{Perez_Lilienfeld:2011}, with more convenient estimators introduced in Ref.~\onlinecite{Cheng_Ceriotti:2014}. Note, however, that in the more general case of free energy difference calculations one can come across a problem that cannot be solved efficiently with either of these two general strategies. For these complicated cases a third option is to run an adiabatic free energy dynamics simulation\cite{Rosso_Tuckerman:2002} and then calculate the free energy difference from the ratio of the resulting probability distributions.\cite{Abrams_Tuckerman:2006,Wu_Yang:2011} Because methods based on finding ratios of probability densities may run into difficulties discussed, for example, in
Ref.~\onlinecite{Kastner_Thiel:2005}, we do not consider them here.

In Ref.~\onlinecite{Karandashev_Vanicek:2017a}, we introduced a method following the first philosophy for eliminating the integration error---in particular, we showed that an effective Monte Carlo procedure for changing the mass reduces the thermodynamic integration error and that a special mass-dependent biasing potential renders the integration error exactly zero. Here, we explore the second strategy, namely the free energy perturbation, or ``direct estimator''
approach;\cite{Perez_Lilienfeld:2011,Cheng_Ceriotti:2014} while direct estimators work best for isotope effects close to unity, they can be applied to larger isotope effects as well by rewriting Eq.~(\ref{eq:IE_definition}) as a product of several smaller isotope effects,\cite{Perez_Lilienfeld:2011,Cheng_Ceriotti:2016} which, incidentally, makes the resulting \textquotedblleft stepwise\textquotedblright\ method
reminiscent of thermodynamic integration. First, we investigate ways to optimize the choice of the smaller isotope effects and the way they are evaluated in order to decrease statistical error of the calculated isotope effect. Second, since the statistical convergence of thermodynamic integration can be improved by changing the mass stochastically during the
simulation, it seems natural to combine this stepwise approach with the procedure for changing mass introduced in Ref.~\onlinecite{Karandashev_Vanicek:2017a}; indeed, testing performance of the resulting combined method on large isotope effects was the main goal of this paper. Such a combination with direct estimators is only possible for a mass sampling procedure that allows finite steps with respect to mass, making the Monte Carlo procedure of Ref.~\onlinecite{Karandashev_Vanicek:2017a}
suitable for the task, but disqualifying standard $\lambda$-dynamics algorithms based on molecular dynamics. Lastly, it is interesting to consider mass-scaled direct estimators used in this work as a \emph{targeted free energy perturbation}\cite{Jarzynski:2002} method, that is a method that uses a physically motivated coordinate mapping (in this case transforming to and from mass-scaled normal modes) to facilitate free energy perturbation calculations. Since all methods discussed in this work rely in some way on the mass-scaled normal mode transformation, their comparison can be rephrased as finding the most efficient way to use a targeted free energy perturbation transformation for a free energy difference calculation. This point will be elaborated further in Sec.~\ref{sec:conclusion}.

To assess the numerical performance of the proposed methodology, we apply it to evaluate isotope effects in an eight-dimensional harmonic model and in full-dimensional methane ($\mathrm{CH}_{4}$) molecule and methanium ($\mathrm{CH}_{5}^{+}$) cation. Methane was chosen because $\mathrm{CH_{4}+D_{2}}$ exchange is an important benchmark reaction for studying catalysis of hydrogen exchange over metals\cite{Osawa_Lee:2010} and metal oxides,\cite{Hargreaves_Taylor:2002} and because the polydeuterated species $\mathrm{CH_{4-x}D_{x}}$ are formed in abundance during the catalyzed reaction. We therefore demonstrate how our new Monte Carlo procedure allows computing not only the large $\mathrm{CD_{4}/CH_{4}}$ isotope effects, but also all $\mathrm{CH_{4-x}D_{x}/CH_{4}}$ isotope effects (x $=1,\ldots,4$) within a single simulation. Methanium was used as an example of a highly fluxional system with very small barriers separating 120 equivalent local minima.\cite{Marx_Parrinello:1995,Marx_Parrinello:1997,Marx_Parrinello:1999} As a consequence, the equilibrium properties of methanium, unlike those of methane, cannot be reliably estimated with the harmonic approximation. We again evaluate all isotope effects of the form $\mathrm{CH_{5-x}D_{x}^{+}/CH_{5}^{+}}$ (x $=1,\ldots,5$).

\section{Theory}

\label{sec:Theory}

In this section, we explain how stochastic change of
mass\cite{Karandashev_Vanicek:2017a} can be combined with the method of direct
estimators\cite{Cheng_Ceriotti:2014} in order to accelerate isotope effect
calculations. In particular we show that for large isotope effects, the
stochastic implementation allows reducing the statistical error while keeping
the integration error zero. We start with a brief overview of the path
integral formalism, thermodynamic integration, and stochastic implementation
of the thermodynamic integration. For more details, see
Ref.~\onlinecite{Karandashev_Vanicek:2017a}, whose notation is followed here.

\subsection{Path integral representation of the partition function}

\label{subsec:path_integrals}

To evaluate the isotope effect~(\ref{eq:IE_definition}) with path integrals,
one first needs a path integral representation of the partition function
$Q=\operatorname{Tr}\exp(-\beta\hat{H})$. For a molecular system consisting of
$N$ atoms with masses $m_{i}$ ($i=1,\ldots,N$) moving in $D\,(=3)$ spatial
dimensions, the partition function can be expressed\cite{Feynman_Hibbs:1965,
Chandler_Wolynes:1981} as the limit $Q=\lim_{P\rightarrow\infty}Q_{P}$, where
$P$ is the number of imaginary time slices (the Trotter number) and
\begin{equation}
Q_{P}=\int d\mathbf{r}\rho(\mathbf{r}) \label{eq:Qr_PI}%
\end{equation}
is the discretized path integral representation of $Q$. {The vector
$\mathbf{r}$ contains all $NDP$ coordinates of all }$N$ {atoms in all }$P$
{slices of the extended configuration space; in particular, $\mathbf{r}%
:=\left(  \mathbf{r}^{(1)},\ldots,\mathbf{r}^{(P)}\right)  $, where
}$\mathbf{r}^{\left(  s\right)  }$, $s=1,\ldots,P$, is a vector containing
$ND$ coordinates of all atoms in slice $s$. The statistical weight
$\rho(\mathbf{r})$ of each path integral configuration is given by%
\begin{equation}
\rho(\mathbf{r})=C\exp\left[  -\beta\Phi(\mathbf{r})\right]  \text{,}%
\end{equation}
with the prefactor%
\begin{equation}
C=\left(  \frac{P}{2\pi\beta\hbar^{2}}\right)  ^{NDP/2}\left(  \prod_{i=1}%
^{N}m_{i}\right)  ^{DP/2}%
\end{equation}
and with an effective potential energy of the classical ring polymer given by
\begin{equation}
\Phi(\mathbf{r})=\frac{P}{2\beta^{2}\hbar^{2}}\sum_{i=1}^{N}m_{i}\sum
_{s=1}^{P}|\mathrm{r}_{i}^{(s)}-\mathrm{r}_{i}^{(s-1)}|^{2}+\frac{1}{P}%
\sum_{s=1}^{P}V(\mathbf{r}^{(s)})\text{,}%
\end{equation}
where $\mathrm{r}_{i}^{(s)}$ denotes the $D$ components of $\mathbf{r}^{(s)}$
corresponding to atom $i$, and $V$ is the potential energy of the original
system. Since the path employed to represent the partition function is a
closed path, we define $\mathbf{r}^{(0)}:=\mathbf{r}^{(P)}$.

Note that $Q_{P}$ is a classical partition function of a ring polymer, a
system in the extended configuration space with $NDP$ classical degrees of
freedom exposed to the effective potential $\Phi(\mathbf{r})$. For $P=1$ the
quantum path integral expression (\ref{eq:Qr_PI}) reduces to the classical
partition function.

\subsection{Thermodynamic integration with respect to mass}

\label{subsec:TI}

A convenient way to evaluate the isotope effect (\ref{eq:IE_definition}) is
based on thermodynamic integration\cite{Kirkwood:1935} with respect to mass.\cite{Vanicek_Aoiz:2005} The isotope change is assumed to be continuous and parametrized by a dimensionless parameter $\lambda\in\lbrack0,1]$, where
$\lambda=0$ corresponds to isotopologue $A$ and $\lambda=1$ to isotopologue $B$. The simplest possible choice for the mass interpolating function is a linear interpolation,\cite{Vanicek_Aoiz:2005,Vanicek_Miller:2007,Zimmermann_Vanicek:2009} but faster convergence is often achieved by interpolating the inverse square
roots of the masses,\cite{Ceriotti_Markland:2013}
\begin{equation}
\frac{1}{\sqrt{m_{i}(\lambda)}}=\left(  1-\lambda\right)  \frac{1}{\sqrt
{m_{i}^{(A)}}}+\lambda\frac{1}{\sqrt{m_{i}^{(B)}}}, \label{eq:m_root_interpol}%
\end{equation}
which is therefore the interpolation we will use in the numerical examples
below. (One can show rigorously that this interpolation is the optimal one in
harmonic systems in the low temperature limit, but has a good behavior at high
temperature and in various other systems as well.)

If $Q(\lambda)$ denotes the partition function of a fictitious system with
interpolated masses $m_{i}(\lambda)$, then the isotope effect
(\ref{eq:IE_definition})\ can be expressed as an exponential of the \textquotedblleft
thermodynamic integral\textquotedblright\
\begin{align}
\frac{Q^{(B)}}{Q^{(A)}}  &  =\exp\left[  \int_{0}^{1}\frac{d\ln Q(\lambda
)}{d\lambda}d\lambda\right] \nonumber\\
&  =\exp\left[  -\beta\int_{0}^{1}\frac{dF(\lambda)}{d\lambda}d\lambda\right]
, \label{eq:introducing_thermodynamic_integration}%
\end{align}
where $F(\lambda)$ is the free energy corresponding to the isotope
change.\ While it is difficult to evaluate either $Q_{P}^{(A)}$ or
$Q_{P}^{(B)}$ with a path integral Monte Carlo method, the logarithmic
derivative $d\ln Q_{P}(\lambda)/d\lambda=[dQ_{P}(\lambda)/d\lambda
]/Q_{P}(\lambda)=-\beta dF_{P}(\lambda)/d\lambda$ is a normalized quantity and
therefore can be computed easily with the Metropolis algorithm with sampling
weight $\rho^{(\lambda)}(\mathbf{r})$ corresponding to the fictitious system
with masses $m_{i}(\lambda)$:%
\[
dF_{P}(\lambda)/d\lambda=\left\langle \left[  dF(\lambda)/d\lambda\right]
_{\text{est}}\right\rangle ^{(\lambda)}.
\]
Here we used a general notation%
\[
\left\langle A_{\text{est}}\right\rangle ^{(\lambda)}:=\frac{\int
d\mathbf{r}A_{\text{est}}(\mathbf{r})\rho^{(\lambda)}(\mathbf{r})}{\int
d\mathbf{r}\rho^{\left(  \lambda\right)  }(\mathbf{r})}%
\]
for a path integral average of an observable $A$, given by averaging the
estimator $A_{\text{est}}$ over an ensemble with weight $\rho^{(\lambda)}$. An
estimator for a quantity $A$ is not unique; different choices of the estimator
result in different statistical behavior. As the analogous centroid virial
estimator for energy\cite{Herman_Berne:1982,Parrinello_Rahman:1984} (obtained
by differentiating with respect to inverse temperature $\beta$), the centroid
virial estimator for the isotope change\cite{Vanicek_Miller:2007} (where the
differentiation is with respect to $\lambda$) has a statistical error that is
independent of $P$. This estimator, used in all our numerical examples, is
given by%
\begin{equation}
\begin{split}
\left[  dF(\lambda)/d\lambda\right]  _{\text{cv}}=&-\sum_{i=1}^{N}\frac
{1}{2m_{i}}\frac{dm_{i}}{d\lambda}\left\{ \frac{D}{\beta}\vphantom{+\frac{1}{P}%
\sum_{s=1}^{P}\left[  (\mathrm{r}_{i}^{(s)}-\mathrm{r}_{i}^{(C)})\cdot
\nabla_{i}V(\mathbf{r}^{(s)})\right]}\right.\\
& \left.+\frac{1}{P}%
\sum_{s=1}^{P}\left[  (\mathrm{r}_{i}^{(s)}-\mathrm{r}_{i}^{(C)})\cdot
\nabla_{i}V(\mathbf{r}^{(s)})\right]  \right\}  , \label{eq:F_cv}%
\end{split}
\end{equation}
where
\begin{equation}
\mathbf{r}^{(C)}:=\frac{1}{P}\sum_{s=1}^{P}\mathbf{r}^{(s)}%
\end{equation}
is the centroid coordinate and $\nabla_{i}$ is the gradient with respect to the coordinates of particle $i $. While in path integral molecular dynamics gradients of $V$ are available and the centroid virial estimator is \textquotedblleft
free,\textquotedblright\ in path integral Monte Carlo implementations only the potential energy itself is required for sampling, and in order to avoid an unnecessary evaluation of forces, one may evaluate the centroid virial estimators by a single finite difference differentiation with respect to
$\lambda$.\cite{Vanicek_Miller:2005,Vanicek_Miller:2007}

To summarize, the calculation of the isotope effect requires running simulations at different values of $\lambda$ and then numerically evaluating the integral in Eq.~(\ref{eq:introducing_thermodynamic_integration}).

\subsection{Stochastic thermodynamic integration}

\label{subsec:stochastic_TI}

Due to the discretization of the thermodynamic integral, the method of
thermodynamic integration necessarily introduces an integration error. While
Ceriotti and Markland\cite{Ceriotti_Markland:2013} reduced the integration
error by optimizing the interpolation functions $m_{i}(\lambda)$, and thus
obtained Eq.~(\ref{eq:m_root_interpol}), Mar\v{s}\'{a}lek and
Tuckerman\cite{Marsalek_Tuckerman:2014} introduced higher-order derivatives of
$Q(\lambda)$ with respect to $\lambda$. Alternatively, we have
shown\cite{Karandashev_Vanicek:2017a} that one can remove the integration error completely by including the $\lambda$ variable as an additional dimension in the Monte Carlo simulation, in a similar manner to the more general $\lambda$-dynamics.\cite{Liu_Berne:1993,Kong_Brooks:1996,Guo_Kong:1998,Bitetti-Putzer_Karplus:2003,Perez_Lilienfeld:2011}%

More precisely, the $\lambda$-interval $[0,1]$ is divided into $J$
subintervals $I_{j}:=[\lambda_{j-1},\lambda_{j}]$, $j=1,\ldots,J$, with
$\lambda_{0}=0<\lambda_{1}<\cdots<\lambda_{J}=1$, and the isotope effect is
then calculated as
\begin{equation}
\frac{Q_{P}^{(B)}}{Q_{P}^{(A)}}=\lim_{J\rightarrow\infty}\exp\left[
-\frac{\beta}{J}\sum_{j=1}^{J}\left\langle \left[  dF(\lambda)/d\lambda
\right]  _{\text{cv}}\right\rangle ^{I_{j}}\right]  ,
\label{eq:STI_expression}%
\end{equation}
where $\langle\cdots\rangle^{I_{j}}$ denotes an average over all path integral
configurations as well as over all $\lambda\in I_{j}$. The simultaneous
stochastic evaluation of the thermodynamic and path integrals permits using a
much larger number of integration steps (i.e., $J$), rendering the integration
error negligible; if the simulation is, in addition, subject to a special
umbrella biasing potential that is a piecewise linear function of $\lambda$,
we have proven that the integration error becomes exactly
zero.\cite{Karandashev_Vanicek:2017a} Our main goal here is evaluating whether
the stochastic change of mass can be combined effectively with another method
for evaluating the isotope effect, namely the method of direct estimators,
which we review now.

\subsection{Free energy perturbation and direct estimators for equilibrium
isotope effects}

\label{subsec:direct_estimators}

\emph{Free energy perturbation}\cite{Zwanzig:1954,Tuckerman:2010}
is an alternative strategy which allows to calculate isotope effects without
introducing an integration
error.\cite{Perez_Lilienfeld:2011,Cheng_Ceriotti:2014} The approach consists
in rewriting Eq.~(\ref{eq:IE_definition}) as
\begin{equation}
\begin{split}
\mathrm{IE}&=\frac{Q_{P}^{(1)}}{Q_{P}^{(0)}}=\frac{\int\rho^{(1)}%
(\mathbf{r})d\mathbf{r}}{\int\rho^{(0)}(\mathbf{r})d\mathbf{r}}\\
&=\frac
{\int[\rho^{(1)}(\mathbf{r})/\rho^{(0)}(\mathbf{r})]\rho^{(0)}(\mathbf{r}%
)d\mathbf{r}}{\int\rho^{(0)}(\mathbf{r})d\mathbf{r}}=\left\langle
\mathcal{Z}_{\mathrm{th}}^{0,1}\right\rangle ^{(0)},
\label{eq:direct_est_introduction}%
\end{split}
\end{equation}
where $\mathcal{Z}_{\mathrm{th}}^{0,1}:=\rho^{(1)}(\mathbf{r})/\rho
^{(0)}(\mathbf{r})$ is the \textquotedblleft thermodynamic direct
estimator,\textquotedblright\cite{Perez_Lilienfeld:2011,Cheng_Ceriotti:2014}
numerically evaluated as
\begin{equation}
\begin{split}
\mathcal{Z}_{\mathrm{th}}^{0,1}=&\left[  \prod_{i=1}^{N}\frac{m_{i}(1)}%
{m_{i}(0)}\right]^{DP/2}\exp\left(  \frac{P}{2\beta\hbar^{2}}\vphantom{\sum_{i=1}%
^{N}\left\{  [m_{i}(0)-m_{i}(1)]\sum_{s=1}^{P}|\mathrm{r}_{i}^{(s)}%
-\mathrm{r}_{i}^{(s-1)}|^{2}\right\} }\right.\\
&\times\left.\sum_{i=1}%
^{N}\left\{  [m_{i}(0)-m_{i}(1)]\sum_{s=1}^{P}|\mathrm{r}_{i}^{(s)}%
-\mathrm{r}_{i}^{(s-1)}|^{2}\right\}  \right)  .
\end{split}\label{eq:thermodynamic_DE}%
\end{equation}
However, a much lower statistical error is obtained by using an alternative,
mass-scaled direct estimator,\cite{Cheng_Ceriotti:2014}
\begin{equation}
\mathcal{Z}_{\mathrm{sc}}^{0,1}=\left[  \prod_{i=1}^{N}\frac{m_{i}(1)}%
{m_{i}(0)}\right]  ^{D/2}\exp\left\{  \frac{\beta}{P}\sum_{s=1}^{P}\left[
V(\mathbf{r}^{(s)})-V(\mathbf{r}_{0,1}^{(s)})\right]  \right\}  ,
\label{eq:mass_scaled_DE}%
\end{equation}
where the scaled coordinates $\mathbf{r}_{\lambda^{\prime},\lambda
^{\prime\prime}}^{(s)}$ are defined as
\begin{equation}
\mathrm{r}_{\lambda^{\prime},\lambda^{\prime\prime},i}^{(s)}:=\mathrm{r}%
_{i}^{(C)}+\sqrt{\frac{m_{i}(\lambda^{\prime})}{m_{i}(\lambda^{\prime\prime}%
)}}(\mathrm{r}_{i}^{(s)}-\mathrm{r}_{i}^{(C)})\text{.}
\label{eq:rescaled_r_defined}%
\end{equation}
The mass-scaled direct estimator can be derived by expressing $Q(1)$ and
$Q(0)$ in Eq.~(\ref{eq:direct_est_introduction}) in terms of mass-scaled
normal mode coordinates $\mathbf{u}$ (see Appendix~A of
Ref.~\onlinecite{Karandashev_Vanicek:2017a} for details). This leads to a
direct estimator $\tilde{\rho}^{(1)}(\mathbf{u})/\tilde{\rho}^{(0)}%
(\mathbf{u})$, which becomes Eq.~(\ref{eq:mass_scaled_DE}) upon transformation
back to standard Cartesian coordinates $\mathbf{r}$. In contrast to the
thermodynamic integration, the method of direct estimators does not introduce
an integration error; the direct estimators, however, are not suitable for
calculating large isotope effects, for which both $\mathcal{Z}_{\mathrm{th}%
}^{0,1}$ and $\mathcal{Z}_{\mathrm{sc}}^{0,1}$ exhibit large statistical
errors because one always uses a sampling weight $\rho^{(0)}(\mathbf{r})$ even
though the natural weight changes from $\rho^{(0)}(\mathbf{r})$ to $\rho
^{(1)}(\mathbf{r})$. Although isotope effect can be evaluated by
running the simulation either with $\rho^{(0)}(\mathbf{r}) $ or $\rho^{(1)}(\mathbf{r})$,
Cheng and Ceriotti noted\cite{Cheng_Ceriotti:2014}
that running the simulation at the heavier isotope leads to smaller
statistical errors of $\mathcal{Z}_{\mathrm{sc}}$; in
Appendix~\ref{app:ODE_err} we prove this property analytically for
harmonic systems. In the process we also encounter cases in which
direct estimators exhibit infinitely large root mean square errors, making
the calculation impossible to converge; however, as shown in
Appendix~\ref{app:ODE_conv}, such situations cannot occur if one runs the
simulation at the lower-mass isotopologue while averaging $\mathcal{Z}%
_{\mathrm{th}}$ or, for convex or bound potentials, at the higher-mass
isotopologue while averaging $\mathcal{Z}_{\mathrm{sc}}$.

\subsection{Stepwise implementation of the direct estimators}

\label{subsec:stepwise_DE}

A simple way to bypass the issue of large statistical errors of direct estimators
for large isotope effects is performing the calculation stepwise by
factoring the large isotope effect into several smaller isotope effects
between virtual
isotopologues.\cite{Perez_Lilienfeld:2011,Cheng_Ceriotti:2016}
For that purpose, one can, as for thermodynamic integration, introduce a set
of $J+1$ intermediate values $\lambda_{j}$ ($j=0,\dots,J$) such that
$\lambda_{0}=0$, $\lambda_{J}=1$, and write
\begin{equation}
\frac{Q(1)}{Q(0)}=\prod_{j=1}^{J}\frac{Q(\lambda_{j})}{Q(\lambda_{j-1})}.
\label{eq:CC_stepwise}%
\end{equation}
For $J$ large enough, $Q(\lambda_{j})/Q(\lambda_{j-1})$ will be sufficiently
close to unity, and hence can be evaluated with direct estimators with a
reasonably small statistical error. It will prove useful to write
Eq.~(\ref{eq:CC_stepwise}), expressed in terms of direct estimator averages,
in a more general manner as
\begin{equation}
\frac{Q_{P}(1)}{Q_{P}(0)}=\prod_{j=1}^{J}\frac{\langle\mathcal{Z}%
_{\mathrm{sc}}^{\overline{\lambda}_{j},\lambda_{j}}\rangle^{(\overline
{\lambda}_{j})}}{\langle\mathcal{Z}_{\mathrm{sc}}^{\overline{\lambda}%
_{j},\lambda_{j-1}}\rangle^{(\overline{\lambda}_{j})}}, \label{eq:DE_stepwise}%
\end{equation}
by using an arbitrary reference value $\overline{\lambda}_{j}$ from the interval $I_{j}=\left[\lambda_{j-1},\lambda_{j}\right]$ as the $\lambda$-value of the sampling weight used in the $j$th factor of the isotope effect; in a broader context of free energy perturbations this approach is known as ``double-wide'' sampling.\cite{Jorgensen_Ravimohan:1985,Jorgensen_Thomas:2008} From now on we will refer to the stepwise evaluation of the isotope effect via Eq.~(\ref{eq:DE_stepwise}) simply as the method of ``direct estimators.'' Choices of $\overline{\lambda}_{j}$ and $\lambda_{j}$ that minimize the statistical error are discussed in Appendix~\ref{app:DE_err}; in general it appears that $\lambda_{j}=j/J$ ($j=0, \ldots,J$) and $\overline{\lambda}_{j}=(j-1/2)/J$ ($j=1, \ldots,J$) are quite close to the optimal values if inverse square root of mass interpolation~(\ref{eq:m_root_interpol}) is used, and therefore were used throughout this work.

\subsection{Combining direct estimators with the stochastic change of mass}

\label{subsec:direct_estimators_in_extended_space}

In the original use of direct estimators,\cite{Cheng_Ceriotti:2014} the
largest isotope effect was rather small, and hence it was possible to evaluate
several isotope effects in a single simulation. In situations where the
isotope effect is large, one should use the generalized, stepwise
expression~(\ref{eq:CC_stepwise}) to compute it, in order to avoid large
statistical errors. In the previous subsection we assumed that each of the
factors contributing to the product~(\ref{eq:CC_stepwise}) is obtained from a
separate simulation, as in Eq.~(\ref{eq:DE_stepwise}).

Here, we propose, instead, to run a single simulation which will explore all
$\overline{\lambda}_{j}$ values at once in a way similar to the stochastic
thermodynamic integration with respect to mass from
Ref.~\onlinecite{Karandashev_Vanicek:2017a} and described briefly in
Subsection~\ref{subsec:stochastic_TI}. There are several reasons why this
should be advantageous.

The first, most obvious reason, is the same as for stochastic thermodynamic
integration from Ref.~\onlinecite{Karandashev_Vanicek:2017a}: given fixed
computational resources, it is much easier to reach ergodicity, and therefore
converge a single Monte Carlo simulation than $J$ separate simulations; in
other words, it is computationally less expensive to obtain a converged
isotope effect if all the factors in the right-hand side of
Eq.~(\ref{eq:DE_stepwise}) are obtained from a single simulation rather than
from $J$ separate simulations.

The second reason is similar to the motivation for the original method of
direct estimators,\cite{Cheng_Ceriotti:2014} in which a single Monte Carlo
simulation suffices to evaluate several different isotope effects; calculating
all factors appearing in Eq.~(\ref{eq:DE_stepwise}) in a single simulation
will indeed enable this at least for \textquotedblleft
sequential\textquotedblright\ isotope effects, such as $\mathrm{CH_{4-x}%
D_{x}/CH_{4}}$ for $x=1,\ldots,4$ or $\mathrm{CH_{5-x}D_{x}%
^{+}/CH_{5}^{+}}$ for $x=1,\ldots,5$ that will be presented below.

Last but not least, the stochastic change of $\lambda$ can lead to a decrease
of statistical error of the calculated isotope effect. Consider $J$ sets of
correlated samples obtained from $J$ independent simulations using the method
of direct estimators and run at different values of $\bar{\lambda}_{j}$;
reshuffling the samples between simulations should make samples inside each
simulation less correlated between each other, thus lowering statistical error
of averages obtained from them. The proposed method can be regarded as such a
\textquotedblleft shuffling,\textquotedblright\ and in this respect it
resembles the parallel tempering or replica exchange Markov chain Monte Carlo
techniques,\cite{Swendsen_Wang:1986,Hukushima_Nemoto:1996,Predescu_Ciobanu:2005}
but for the latter approaches the value of $J$ that can be used in practice
will depend on the number of simulation replicas one can run simultaneously.

With this motivation in mind, it is easy to see that to change $\lambda$ value
between different $\overline{\lambda}_{j}$ values one can use the same
procedure as that described in Ref.~\onlinecite{Karandashev_Vanicek:2017a},
the only difference being that the trial $\lambda$ value should be restricted
to a discrete set of values $\{\overline{\lambda}_{j}\}$, $j=1,\ldots,J$; some
more tedious details are left for Appendix~\ref{app:SDE_details}. The overall
isotope effect is again obtained from Eq.~(\ref{eq:DE_stepwise}).

Combining stochastic change of mass with the thermodynamic integration or stepwise direct estimators also makes it possible to calculate several isotope effects at once more efficiently. This result is discussed properly in Supplementary Material using deuterization of methane and methanium as examples.

\section{Numerical examples}

\label{sec:Applications}

To test the stepwise and stochastic implementations of the direct estimators, we applied them to a model $8$-dimensional harmonic system and to several isotopologues of methane and methanium. In addition we compared the results of the new approaches with results of the thermodynamic integration with either deterministic or stochastic change of mass, and with the original method of direct
estimators.\cite{Cheng_Ceriotti:2014} From now on, for brevity we will refer
to the five methods as follows: thermodynamic integration with respect to mass
(Subsection~\ref{subsec:TI}) will be simply referred to as \textquotedblleft
thermodynamic integration\textquotedblright\ (TI), thermodynamic integration
with stochastic change of mass (Subsection~\ref{subsec:stochastic_TI}) as
\textquotedblleft stochastic thermodynamic integration\textquotedblright%
\ (STI), Cheng and Ceriotti's original method of direct estimators
(Subsection~\ref{subsec:direct_estimators}) as \textquotedblleft original
direct estimators\textquotedblright\ (ODE), stepwise application of direct
estimators (Subsection~\ref{subsec:stepwise_DE}) as \textquotedblleft direct
estimators\textquotedblright\ (DE), and stepwise application of direct
estimators with stochastic change of mass (Subsection~\ref{subsec:direct_estimators_in_extended_space}) as \textquotedblleft stochastic direct estimators\textquotedblright\ (SDE).

\subsection{Computational details}

\label{subsec:global_comp_details}

The calculations presented in this work were done with parameters mostly identical to those used in Ref.~\onlinecite{Karandashev_Vanicek:2017a}. The DE and ODE calculations were done with the same number of different Monte Carlo steps and frequency of estimators' evaluation as TI, while the SDE calculations employed the same number of different Monte Carlo steps and frequency of estimators' evaluation as STI. In SDE calculations, we additionally modified simple $\lambda$-moves used in STI calculations according to a prescription detailed in Appendix~\ref{app:SDE_details}. The way the $\lambda$ interval was divided into subintervals was also the same as in Sec.~III of Ref.~\onlinecite{Karandashev_Vanicek:2017a}.

As discussed in Sec.~\ref{subsec:TI}, results obtained with TI contain integration error, which was estimated by comparing the calculated isotope effects with the exact analytical\cite{Schweizer_Wolynes:1981} values for a
harmonic system with a finite Trotter number $P$ and with the result of SDE for methane. (Recall that ODE, DE, and SDE have no integration error by definition, while the absence of integration error in STI was proven in Ref.~\onlinecite{Karandashev_Vanicek:2017a}.) Statistical errors were evaluated with the \textquotedblleft block-averaging\textquotedblright\ method\cite{Flyvbjerg_Petersen:1989} for correlated samples the same way as in Sec.~III of Ref.~\onlinecite{Karandashev_Vanicek:2017a}. Last but not
least, the path integral discretization error for the harmonic system was estimated from a known analytical expression for finite $P$,\cite{Schweizer_Wolynes:1981} while for the $\mathrm{CD_{4}/CH_{4}}$ and $\mathrm{CD_{5}^{+}/CH_{5}^{+}}$ isotope effects it was estimated using a novel procedure presented in Appendix~\ref{app:discr_error}.

\subsection{\label{subsec:SDE_harmonic}Isotope effects in a harmonic model}

\begin{figure*}
\centering\includegraphics[width=0.75\textwidth]{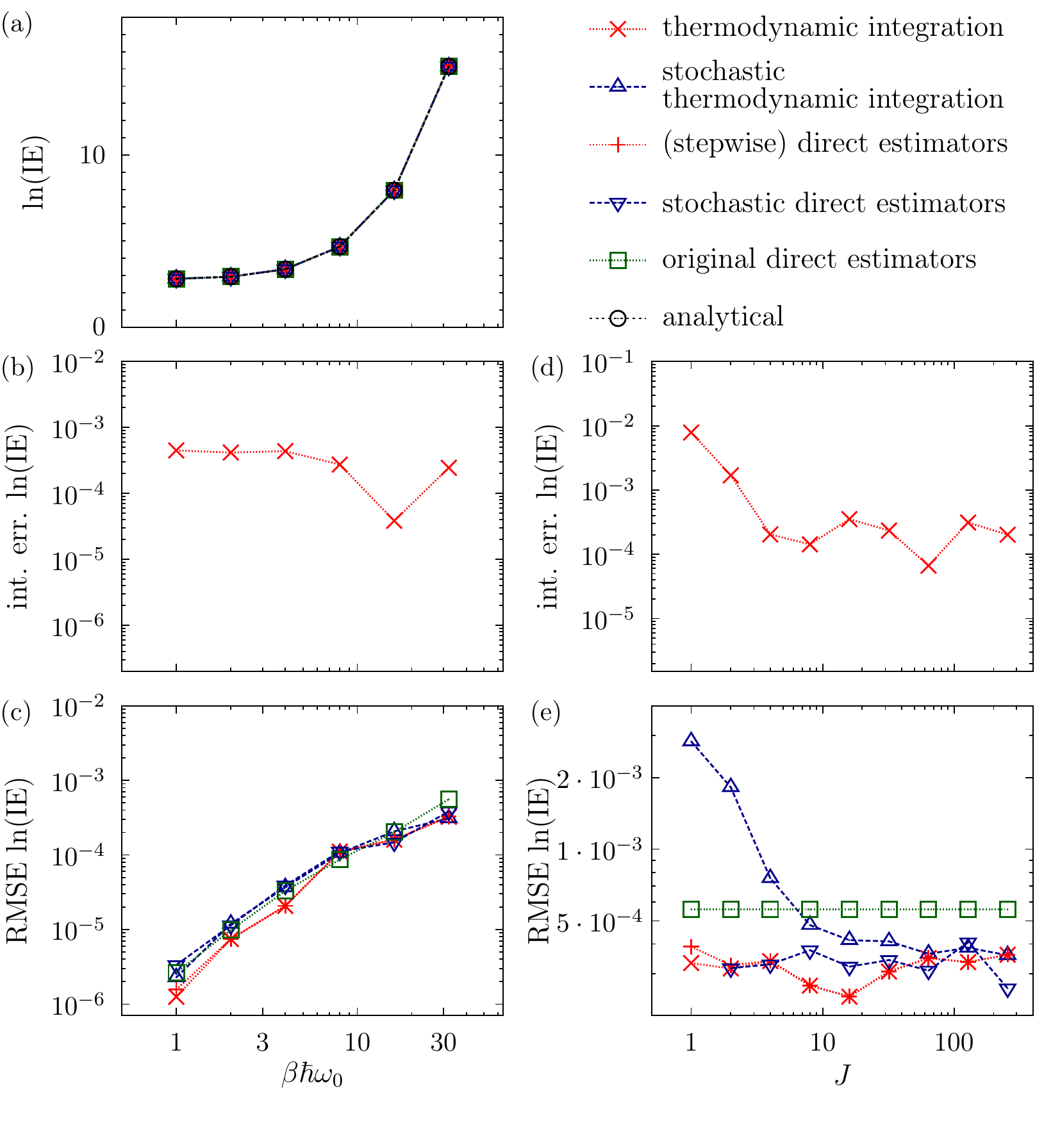}
\caption{\label{fig:SDE_ho_num_performance}Isotope effect calculations in an eight-dimensional harmonic model from Ref.~\onlinecite{Karandashev_Vanicek:2017a}. Results of thermodynamic integration (TI), stochastic thermodynamic integration (STI),
stepwise direct estimators (DE), stochastic direct estimators (SDE), and original direct estimator (ODE) approach
from Ref.~\onlinecite{Cheng_Ceriotti:2014} are compared
with exact analytical values (for the same finite Trotter number $P$). The proposed method is
SDE, while DE already provide some improvement over ODE.
Panels (a)-(c) show the temperature dependence of
(a) the isotope effect,
(b) its integration errors for TI, and
(c) its statistical root mean square errors (RMSEs).
Panels (d)-(e) display the dependence of integration errors (for TI) and RMSEs on the number $J$ of
integration subintervals at a temperature given by $ \beta\hbar\omega_{0}=32$.}
\end{figure*}

Numerical results are presented in Fig.~\ref{fig:SDE_ho_num_performance} and
correspond to an $8$-dimensional harmonic system with parameters identical to
those used in Ref.~\onlinecite{Karandashev_Vanicek:2017a}. Panel~(a) of
Fig.~\ref{fig:SDE_ho_num_performance} shows that analytical values of the
isotope effect (at a finite value of $P$) are reproduced accurately by all
five methods for several values of $\beta\hbar\omega_{0}$, confirming, in
particular, that the proposed stepwise and stochastic implementations of the
direct estimators are correct.

Panel~(b) displays the dependence of the thermodynamic integration error on
temperature. (Recall that the integration error is zero for STI, ODE, DE, and
SDE by construction). The figure is a reminder of the fact that, despite the
improved mass interpolation scheme~(\ref{eq:m_root_interpol}), TI is the only
of the five presented methods that exhibits a significant integration error,
especially noticeable at higher temperatures since the mass interpolating
function Eq.~(\ref{eq:m_root_interpol}) was designed to be most effective in
the deep quantum regime.

Panel~(c) of Fig.~\ref{fig:SDE_ho_num_performance} compares statistical errors of the five methods considered. The first evident trend is that at lower temperatures ODE exhibit a larger statistical error compared to the
other methods; it illustrates the need to use the stepwise (DE) or stochastic
stepwise (SDE) variants instead in such cases. Secondly, similar statistical
errors are exhibited by TI and DE, as well as by STI and SDE. This can be
explained by noticing that in the limit of large $J$ TI becomes equivalent to
DE, while STI becomes equivalent to SDE; to see this one can recall that
$[dF(\lambda)/d\lambda]_{\mathrm{est}}$ is related to the derivative of
$\mathcal{Z}_{\mathrm{sc}}^{\lambda^{\prime},\lambda^{\prime\prime}}$ with
respect to $\lambda^{\prime\prime}$ and compare Eq.~(\ref{eq:DE_stepwise}),
Eq.~(\ref{eq:STI_expression}), and the midpoint rule used for thermodynamic
integration in this work [also see Eq.~(32) of
Ref.~\onlinecite{Karandashev_Vanicek:2017a}]. It is therefore reasonable to
expect that for large values of $J$ or small isotope effects statistical
errors of TI and DE or STI and SDE will be quite close; here $J=8$ is
apparently already sufficient to enforce this tendency over a wide temperature range.

Panel~(d) of Fig.~\ref{fig:SDE_ho_num_performance} displays integration errors
of TI as a function of the number $J$ of $\lambda$ intervals. Clearly, for TI,
the integration error approaches zero only in the $J\rightarrow\infty$ limit,
which is, exceptionally, attainable in this particular case since the Monte Carlo
procedure produces uncorrelated samples. Note, however, that in most practical
calculations, this is impossible and the limit cannot be reached.

Finally, panel~(e) of Fig.~\ref{fig:SDE_ho_num_performance} shows the
dependence of statistical errors on $J$, demonstrating that the statistical
errors of the four methods approach their limiting values for $J\rightarrow
\infty$. As already mentioned earlier, similar statistical errors for TI and
DE, or for STI and SDE in the $J\rightarrow\infty$ limit are expected. Here,
however, the tendency is already observed at a surprisingly small value of $J=1$ for TI and DE. For STI and SDE it is observed after $J=64$.

\subsection{Deuteration of methane}

\label{subsec:SDE_methane_deuterization}

\subsubsection{Computational details}

\label{subsubsec:SDE_comp_det_methane}

As mentioned above, calculation parameters for DE, SDE, and ODE
are identical to those from Ref.~\onlinecite{Karandashev_Vanicek:2017a}, with
a few modifications described in Subsection~\ref{subsec:global_comp_details};
the TI and STI results were taken from
Ref.~\onlinecite{Karandashev_Vanicek:2017a}. Here we only add that the
computational time spent on TI, STI, SDE, DE, and ODE calculations was
approximately the same.

For each temperature we also ran calculations at $\lambda=0$ and $\lambda=1$
to estimate the path integral discretization error with the method explained
in Appendix~\ref{app:discr_error} which uses estimator $\mathcal{W}_{2}$ for $Q_{2P}/Q_{P}$. These simulations were $10^{7}$ Monte
Carlo steps long, $14\%$ were whole-chain moves and $86\%$ were multi-slice
moves performed on one sixth of the chain with the staging
algorithm\cite{Sprik_Chandler:1985,Sprik_Chandler:1985_1} (this guaranteed
that approximately the same computer time was spent on either of the two types
of moves). $\mathcal{W}_{2}$ was evaluated only after each ten Monte
Carlo steps to decrease the calculation cost.

\subsubsection{Results and discussion}

\begin{figure}
\centering\includegraphics[width=0.375\textwidth, trim={0 0 8.4cm 0}]{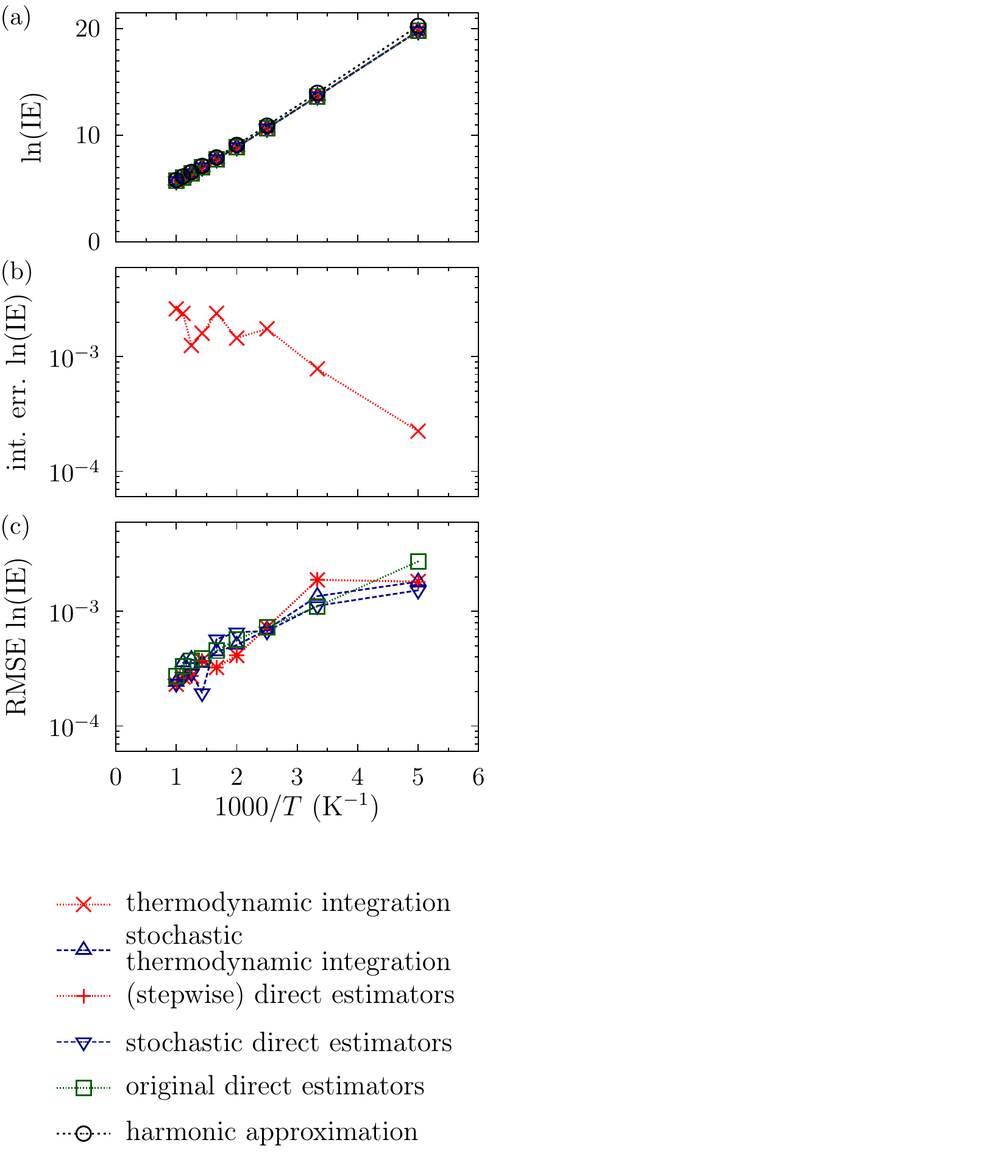}
\caption{\label{fig:SDE_ch4_num_performance} $\mathrm{CD}_{4}/\mathrm{CH}_{4} $ isotope effect (IE) computed with several methods. Panels (a)-(c) show the temperature dependence of (a) the isotope effect, (b) its integration errors for thermodynamic integration, and (c) its statistical root mean square errors (RMSEs).}
\end{figure}

The results of the calculations of the $\mathrm{CD}_{4}/\mathrm{CH}_{4}$ isotope effect are presented in
Fig.~\ref{fig:SDE_ch4_num_performance}, with TI and STI results from
Ref.~\onlinecite{Karandashev_Vanicek:2017a} plotted for comparison. Panel~(a)
shows that the isotope effects calculated with the five different methods
described in Sec.~\ref{sec:Theory} agree, confirming that SDE were implemented
correctly, even though TI does exhibit a small
integration error [see panel~(b)]. Panel~(c) of Fig.~\ref{fig:SDE_ch4_num_performance} shows
that STI, SDE, TI, and DE exhibit quite similar statistical errors, while the ODE,
quite surprisingly, exhibit statistical errors quite close to
stepwise approaches for all but the lowest temperature. Several reasons why ODE
seem to perform well for a rather large range of isotope effect values are
analyzed in Appendix~\ref{app:ODE_conv}.

For reference, the SDE, DE, and ODE values plotted in Fig.~\ref{fig:SDE_ch4_num_performance} are listed in Table~\ref{tab:lnIE_methane_values_and_disc_error} together with their discretization errors estimated by the method described in Appendix~\ref{app:discr_error}. Note that the discretization error only depends on $P$ and not on the method used for the isotope effect calculation, and that our method for estimating this discretization error exhibits favorable statistical behavior.

\begin{figure}
\centering\includegraphics[width=0.375\textwidth, trim={0 0 8.4cm 0}]{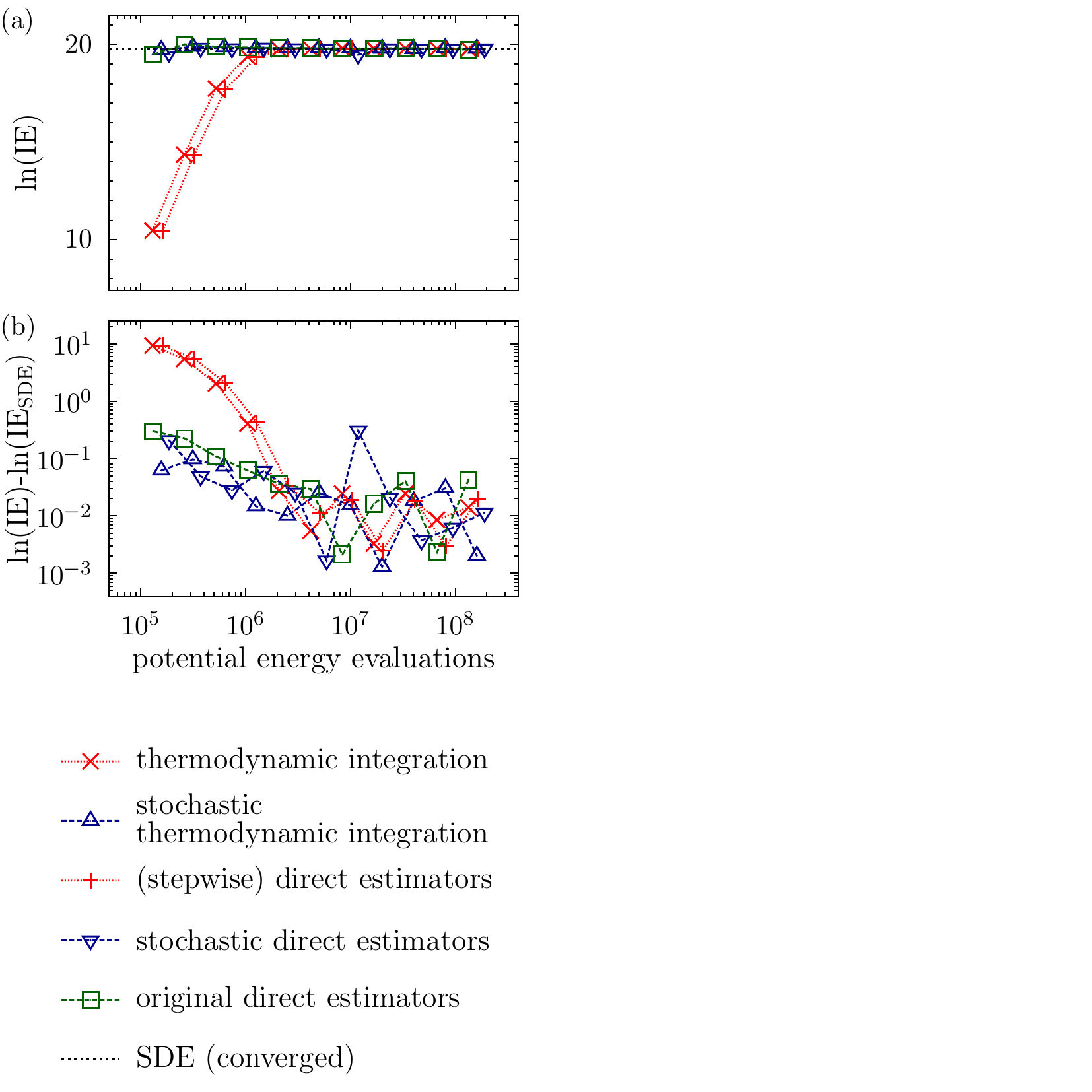}
\caption{\label{fig:SDE_ch4_nonergodicity} Impact of nonergodicity appearing in shorter calculations of the $ \mathrm{CD}_{4}/\mathrm{CH}_{4}$ isotope effect
(IE) at $T=200$ K. Panel
(a) presents the convergence of the IE as a function of the number of potential energy evaluations needed for calculations with different simulation lengths, while panel (b) shows the corresponding error
of the IE (in logarithmic scale) relative to a converged stochastic direct estimators (SDE) result. The horizontal line in panel (a) labeled ``SDE (converged)'' is the converged SDE result
$\ln(\text{IE}_{\mathrm{SDE}})=19.785$ from Fig.~\ref{fig:SDE_ch4_num_performance} and Table~\ref{tab:lnIE_methane_values_and_disc_error}; the same value was used as a reference in panel (b).}
\end{figure}

As mentioned in Ref.~\onlinecite{Karandashev_Vanicek:2017a}, benefits of
stochastically changing mass become most apparent when computational resources
are limited. We therefore compared isotope effects obtained with TI, DE, STI, SDE, and ODE using simulations of increasing length, starting with very short, and hence nonergodic simulations. The results of these calculations, performed according to a prescription detailed in Ref.~\onlinecite{Karandashev_Vanicek:2017a}, are
plotted in Fig.~\ref{fig:SDE_ch4_nonergodicity}. The general tendency is the same as was observed in Ref.~\onlinecite{Karandashev_Vanicek:2017a}: approaches that require several simulations to evaluate the isotope effect (TI and DE) require more Monte Carlo steps in total to achieve converged results than those requiring only one simulation (STI, SDE, and ODE). This is due to a certain time needed for a simulation to adequately explore the $ \mathbf{r}$ space; while TI and DE need to take the corresponding minimal number of Monte Carlo steps for $ J$ simulations, STI, SDE, and ODE need to do it only for one simulation. For STI and SDE this benefit is slightly counteracted by introducing an extra dimension $ \lambda$ to be explored by the simulation, but we have found our Monte Carlo procedure tends to explore this dimension much faster than $ \mathbf{r}$ in realistic simulations; a model example, where this is not the case, is provided in Sec.~II of Supplementary Material.

\subsection{Deuteration of methanium}

\label{subsec:methanium_deuteration}

In this subsection we evaluate the $\mathrm{CD}_{5}^{+}/\mathrm{CH}_{5}^{+}$ isotope effect using the same techniques as for methane in Subsec.~\ref{subsec:SDE_methane_deuterization} and Ref.~\onlinecite{Karandashev_Vanicek:2017a}, including TI, DE, STI, SDE, ODE, and harmonic approximation; the only difference of the calculation parameters was that $J=5 $ was employed for mass discretization. Potential energy surface from Ref.~\onlinecite{Jin_Bowman:2006} was used during the simulations.

\begin{figure}
\centering\includegraphics[width=0.375\textwidth, trim={0 0 8.4cm 0}]{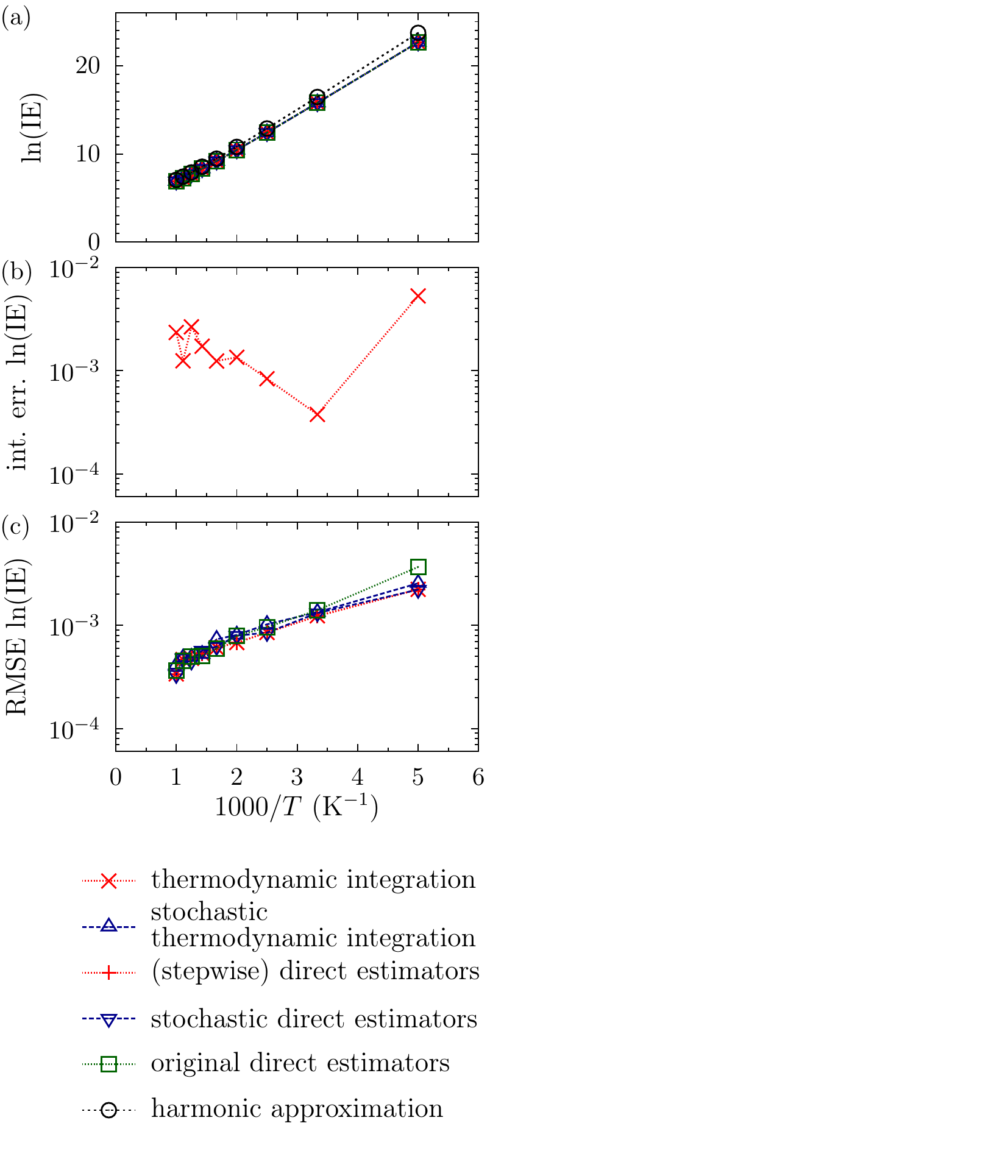}
\caption{\label{fig:ch5_plus_num_performance}$\mathrm{CD}_{5}^{+}/\mathrm{CH}_{5}^{+} $ isotope effect (IE) computed with  several methods. For details, see caption of Figure~\ref{fig:SDE_ch4_num_performance}.}
\end{figure}

The resulting isotope effects are plotted in Fig.~\ref{fig:ch5_plus_num_performance} along with their root mean square errors and integration error for TI. As seen in panel~(a), the four path integral methods that do not exhibit integration error (DE, SDE, STI, and ODE) agree with each other up to statistical error, while TI exhibits a noticeable, but rather small, integration error plotted in panel~(b). Panel~(c) shows root mean square errors of the isotope effect obtained with different methods. As was the case for methane, TI, DE, STI, and SDE exhibit similar statistical errors in the entire temperature range. ODE exhibit statistical errors similar to the other four methods for all temperatures but the lowest one, for which using the stepwise approach decreases the statistical error by approximately $ 40\%$; an analysis of why ODE perform well for isotope effects of such a large magnitude is presented in Appendix~\ref{app:ODE_conv}. Isotope effect values obtained with the harmonic approximation are also plotted in panel~(c) and they agree surprisingly well with exact path integral methods in this case, probably because the greatest contributions to an isotope effect tend to arise from more rigid molecular degrees of freedom (e.g., movement of the carbon atom towards one of the hydrogens), which are largely harmonic.\cite{Huang_Nesbitt:2006,Ivanov_Schlemmer:2010,Ivanov_Marx:2013}   Note that the harmonic approximation was obtained by expanding the potential energy about one of the 120 equivalent potential energy minima of $ \mathrm{CH}_{5}^{+}$. For reference, all calculated isotope effects along with the estimated discretization errors are presented in Table~\ref{tab:lnIE_methanium_values_and_disc_error}.

\begin{figure}
\centering\includegraphics[width=0.375\textwidth, trim={0 0 8.4cm 0}]{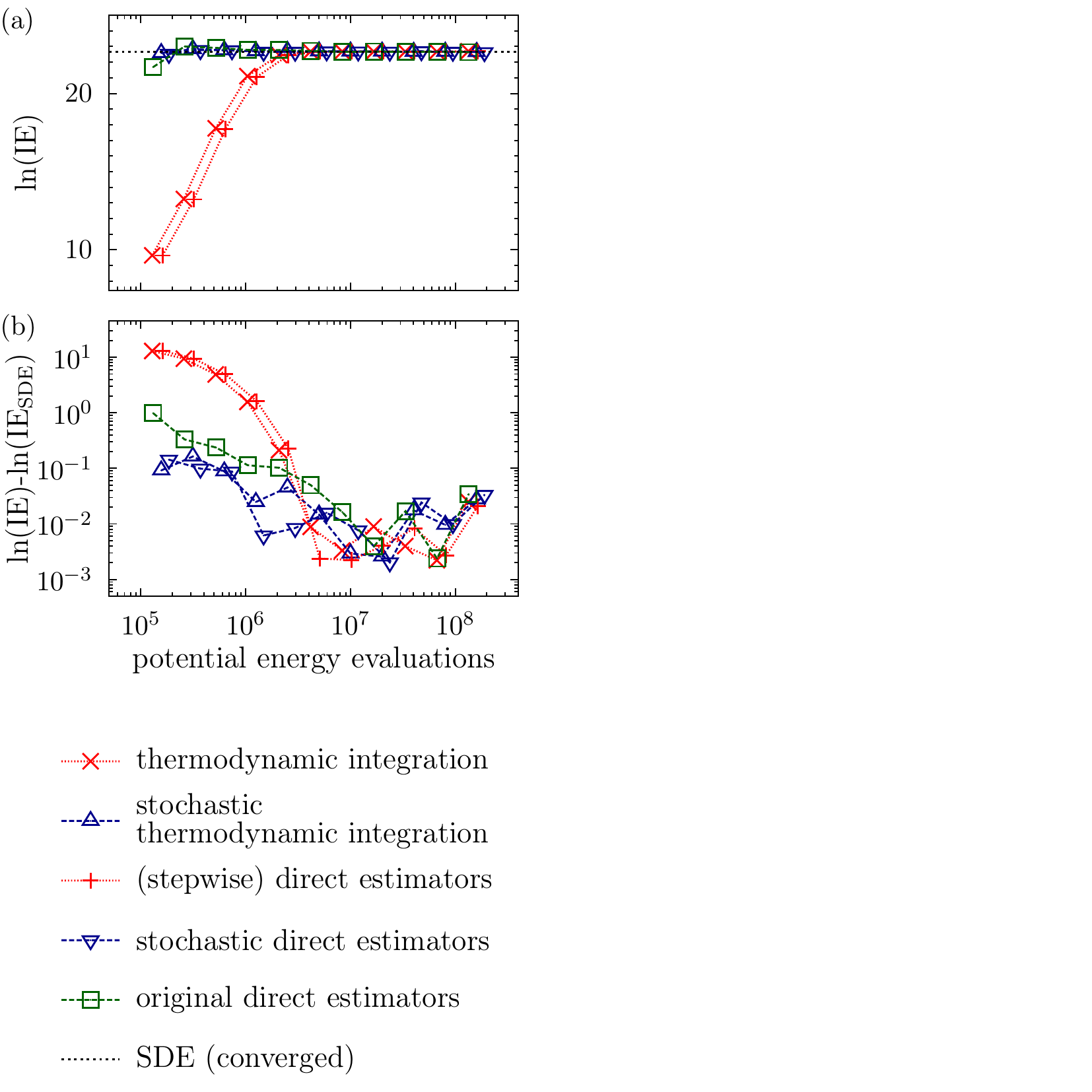}
\caption{\label{fig:SDE_ch5_plus_nonergodicity}Impact of nonergodicity appearing in shorter calculations of the $ \mathrm{CD}_{5}^{+}/\mathrm{CH}_{5}^{+}$ isotope effect (IE) at $T=200$ K. For details, see the caption of Figure~\ref{fig:SDE_ch4_nonergodicity}. The horizontal line in panel (a) labeled ``SDE (converged)'' is the converged SDE result $\ln(\text{IE}_{\mathrm{SDE}})=22.674$ from Fig.~\ref{fig:ch5_plus_num_performance} and Table~\ref{tab:lnIE_methanium_values_and_disc_error};
the same value was used as a reference in panel (b).}
\end{figure}

As in Subsec.~\ref{subsec:SDE_methane_deuterization}, we compared isotope effects obtained with TI, DE, STI, SDE, and ODE using simulations of increasing length, starting with very short, and hence nonergodic simulations. The results of these calculations, performed according to a prescription detailed in Ref.~\onlinecite{Karandashev_Vanicek:2017a}, are plotted in Fig.~\ref{fig:SDE_ch5_plus_nonergodicity}. The general tendency is the same as was observed in Ref.~\onlinecite{Karandashev_Vanicek:2017a} and in the case of methane in Subsec.~\ref{subsec:SDE_methane_deuterization}: approaches that require several simulations to evaluate the isotope effect (TI and DE) require more Monte Carlo steps in total to achieve converged results than those requiring only one simulation (STI, SDE, and ODE).

\begin{table*}
\caption{Values of the $ \mathrm{CD}_{4}/\mathrm{CH}_{4} $ isotope effect (IE) obtained with
direct estimators (DE), ``stochastic direct estimators'' (SDE), and ``original direct estimators'' (ODE)
(for other values seen in Fig.~\ref{fig:SDE_ch4_num_performance}, see Ref.~\onlinecite{Karandashev_Vanicek:2017a}).
The proposed methodology is SDE.
The discretization error is defined as $ \mathrm{IE}_{P}/\mathrm{IE}-1$
and is estimated with the procedure described in Appendix~\ref{app:discr_error}.}
\label{tab:lnIE_methane_values_and_disc_error}\begin{ruledtabular}
\begin{tabular}{cccccc}
\multirow{2}{*}{$ T (\text{K}) $} & \multirow{2}{*}{$ P $} & \multicolumn{3}{c}{ln(IE) ($\mathrm{CD_{4}/CH_{4}}$) with statistical error} &
Estimate of \tabularnewline
\cline{3-5}
&  & DE & SDE & ODE & discretization error ($ 10^{-3}$)\tabularnewline
\hline
\pz200 & 360 & $ 19.786\pm 0.002$ & $19.785\pm 0.002 $
& $19.790\pm 0.003$& $-5.1\pm 0.3$\tabularnewline
\pz300 & 226 & $ 13.672\pm 0.002$ & $ 13.672\pm 0.001$ &
$13.673\pm 0.001$ & $-3.7\pm 0.2$\tabularnewline
\pz400 & 158 & $ 10.666\pm 0.001$ &
$ 10.667\pm 0.001$ & $ 10.666\pm 0.001$
& $-3.2\pm 0.2$\tabularnewline
\pz500 & 118 & $\pz8.910\pm 0.001$ &
$\pz8.910\pm 0.001$ & $\pz8.909 \pm 0.001$
& $-3.0\pm 0.2$\tabularnewline
\pz600 & \pz90 & $ \pz7.780\pm 0.001$ &
$ \pz7.780\pm 0.001 $ & $ \pz7.780 \pm 0.001$
& $-3.0\pm 0.2$\tabularnewline
\pz700 & \pz72  & $\pz7.006\pm 0.001$ &
$\pz7.005\pm 0.001 $ & $\pz7.007\pm 0.001$
& $-2.8\pm 0.2$\tabularnewline
\pz800 & \pz58  & $\pz6.451 \pm 0.001$ &
$\pz6.450\pm 0.001$ & $\pz6.451 \pm 0.001$
& $-2.8\pm 0.2$\tabularnewline
\pz900 & \pz46  & $\pz6.039\pm 0.001$ &
$\pz6.039\pm 0.001$ & $\pz6.038 \pm 0.001$
& $-3.4\pm 0.2$\tabularnewline
1000 & \pz36  & $\pz5.725\pm 0.001$ &
$\pz5.726\pm 0.001 $ & $\pz5.725 \pm 0.001$
& $-3.9\pm 0.2$\tabularnewline
\end{tabular}
\end{ruledtabular}
\end{table*}

\begin{table*}
\caption{Values of the $ \mathrm{CD}_{5}^{+}/\mathrm{CH}_{5}^{+} $ isotope effect (IE) obtained with several different methods. For details, see the caption of Table~\ref{tab:lnIE_methane_values_and_disc_error}. The values of $ P$ for a given temperature were the same as those for methane listed in Table~\ref{tab:lnIE_methane_values_and_disc_error}.
}\label{tab:lnIE_methanium_values_and_disc_error} \begin{ruledtabular}
\begin{tabular}{cccccccc}
\multirow{3}{*}{$ T (\text{K}) $} & \multicolumn{6}{c}{ln(IE) ($\mathrm{CD_{5}^{+}/CD_{5}^{+}}$)} &
\multirow{3}{*}{\begin{tabular}{c}
Estimate of \\ discretization \\ error ($ 10^{-3}$)\end{tabular}} \tabularnewline
\cline{2-7}
& \multicolumn{5}{c}{Path integral values with statistical errors} & \multirow{2}{*}{\begin{tabular}{c}
harmonic \\ approximation\end{tabular}}
\tabularnewline
\cline{2-6}
& TI & DE & STI & SDE & ODE & \tabularnewline
\hline
\pz200 & $22.669\pm 0.003$ & $ 22.669\pm0.003$ & $ 22.671\pm0.003$ & $ 22.674\pm0.003$ & $ 22.665\pm0.004$ & $ 23.727$& $ -5.0\pm 0.3$ \tabularnewline
\pz300 & $ 15.789\pm0.002$ & $ 15.790\pm 0.002$ & $ 15.790\pm0.002$ & $ 15.789\pm0.002$ & $ 15.790\pm0.002$ & $ 16.446$& $ -3.9\pm 0.3$ \tabularnewline
\pz400 & $ 12.411\pm0.001$ & $ 12.412\pm0.001$ & $ 12.410\pm0.001$& $ 12.410\pm0.001$ & $ 12.413\pm0.001$& $ 12.875$ & $ -3.5\pm 0.3$ \tabularnewline
\pz500 & $ 10.443\pm0.001$ & $ 10.444\pm0.001$ & $ 10.445\pm0.001$ & $ 10.444\pm0.001$ & $ 10.443\pm0.001$& $ 10.796$ & $ -3.1\pm 0.2$ \tabularnewline
\pz600 & $ \pz9.181\pm0.001$ & $ \pz9.183\pm0.001$ & $ \pz9.183\pm0.001$ & $ \pz9.182\pm0.001$ & $ \pz9.183\pm0.001$& $ \pz9.461$& $ -3.0\pm 0.2$ \tabularnewline
\pz700 & $ \pz8.319\pm0.001$ & $ \pz8.321\pm0.001$ & $ \pz8.321\pm0.001$ & $ \pz8.321\pm0.001$& $ \pz8.321\pm0.001$ & $ \pz8.549$& $ -2.7\pm 0.2$ \tabularnewline
\pz800 & $ \pz7.703\pm0.001$ & $ \pz7.705\pm0.001$ & $ \pz7.705\pm0.001$ & $ \pz7.706\pm0.001$ & $ \pz7.704\pm0.001$& $ \pz7.896$& $ -3.1\pm 0.2$ \tabularnewline
\pz900 & $ \pz7.248\pm0.001$ & $ \pz7.250\pm0.001$& $ \pz7.249\pm0.001$& $ \pz7.249\pm0.001$ & $ \pz7.250\pm0.001$ & $ \pz7.414$& $ -3.7\pm 0.2$ \tabularnewline
1000 & $ \pz6.902\pm0.001$ & $ \pz6.904\pm0.001$ & $ \pz6.905\pm0.001$ & $ \pz6.905\pm0.001$& $ \pz6.905\pm0.001$ & $ \pz7.048$& $ -4.1\pm 0.2$ \tabularnewline
\end{tabular}
\end{ruledtabular}
\end{table*}

\section{Conclusion}
\label{sec:conclusion}

In this paper, we continued our program of accelerating the calculations of equilibrium isotope effects. Starting from a basic idea of thermodynamic integration with respect to mass,\cite{Vanicek_Aoiz:2005} one can 1)\ accelerate statistical convergence by using centroid virial estimators,
which, in the case of path integral Monte Carlo avoid evaluating forces by employing a finite-difference derivative
trick,\cite{Vanicek_Miller:2007,Zimmermann_Vanicek:2009} 2) accelerate convergence to the quantum limit by employing higher-order factorizations of the path integral,\cite{Buchowiecki_Vanicek:2013} and 3) reduce or remove the
error of thermodynamic integration, which was the focus of the present work.

To remove the thermodynamic integration error,
we investigated optimal ways to use direct estimators\cite{Perez_Lilienfeld:2011,Cheng_Ceriotti:2014} for larger isotope effects. First, we introduced a near optimal way to discretize the larger isotope effect into smaller ones, and showed that the statistical error of the resulting method was quite close
to the corresponding error exhibited by thermodynamic integration. Second, we showed how
statistical convergence of the method could be improved by combining it with the stochastic change of mass. We referred to the resulting method as the ``stochastic direct estimators.''

We also observed that simple, original application of the mass-scaled direct estimators proposed in Ref.~\onlinecite{Cheng_Ceriotti:2014} performed well for isotope effects of surprisingly large magnitudes. These results, combined with our discussion of the method's convergence and divergence in Appendix~\ref{app:ODE_conv}, indicate that the applicability of the direct estimators could be broader than what had been reported before. In the end, the approach of Ref.~\onlinecite{Cheng_Ceriotti:2014} can be quite efficient for isotope effects much larger than unity, especially if the system in question allows to capitalize on atom interchangeability. Yet, for the largest isotope effects using stochastic direct estimators is still preferable, as indicated by our results for the deuterization of methane and methanium at lower temperatures.

The way we used mass-scaled normal modes to define mass-scaled $ \lambda$-moves can be extended to any mapping used in targeted free energy perturbation\cite{Jarzynski:2002} provided the mapping is non-local. In this context our results compare using the same physically-motivated mapping either to perform a free energy perturbation calculation (original direct estimators) or to define an analogue of our Monte Carlo procedure and combine it with thermodynamic integration or stepwise free energy perturbation. The quality of the mapping affects statistical errors of the first approach through dispersion of the free energy perturbation estimator; in the second approach estimator properties are less affected, but an inefficient procedure for sampling $ \lambda$ can increase correlation length of the samples. Our results for methane and methanium at lower temperatures illustrate how the second approach is less affected by the quality of the mapping.

Finally, let us mention that stochastic direct estimators can be combined with the
Takahashi-Imada or Suzuki fourth-order
factorizations\cite{Takahashi_Imada:1984,Suzuki:1995,Chin:1997,Jang_Voth:2001}
of the Boltzmann operator, which would allow lowering the path integral
discretization error of the computed isotope effect for a given Trotter number
$P$, and hence a faster convergence to the quantum
limit.\cite{Perez_Tuckerman:2011,Buchowiecki_Vanicek:2013,Marsalek_Tuckerman:2014,
Karandashev_Vanicek:2015}

\section*{Supplementary material}

Section~I of Supplementary Material describes an alternative procedure, in which several isotope effects are evaluated simultaneously. Examples include four isotope effects of the form $\mathrm{CH_{4-x}D_{x}/CH_{4}}$ (x$=1,\ldots,4$) and five isotope effects $\mathrm{CH_{5-x}D_{x}^{+}/CH_{5}^{+}}$ ($\mathrm{x}=1,\ldots,5$). Section~II analyzes effects of nonergodicity on isotope effect calculations in the harmonic model.

\begin{acknowledgments}
This research was supported by the Swiss National Science Foundation with Grant No. 200020\_150098 as well as within the National Center for Competence in Research \textquotedblleft Molecular Ultrafast Science and Technology\textquotedblright\ (NCCR MUST), and by the EPFL.
\end{acknowledgments}

\bibliographystyle{aipnum4-1}

\appendix

\section{\label{app:ODE_err}Statistical errors, convergence, and divergence of direct estimators for
an isotope effect in a harmonic system}

In this appendix we derive $P\rightarrow\infty$ limits of root mean square errors (RMSEs)\ of direct estimators $\mathcal{Z}_{\mathrm{th}}$ and $\mathcal{Z}_{\mathrm{sc}}$ for a harmonic system. For simplicity, the system is defined as $N$ particles with masses $m_{i}$ moving in one-dimensional Cartesian space, the corresponding oscillator frequencies being $\omega_{i}$. We are mainly interested in the relative statistical error of $A$, that is $\Delta A/A$, where
\begin{equation}
\Delta A=\sqrt{\langle A^{2}\rangle-\langle A\rangle^{2}}%
\end{equation}
is the RMSE of $A$ and $\langle X\rangle$ denotes an average over all
samples of $X$. To simplify the analytical derivation, instead of $\Delta A/A$,
we will evaluate the quantity
\begin{equation}
\left(  \frac{\Delta A}{A}\right)  ^{2}+1=\frac{\langle A^{2}\rangle}{\langle
A\rangle^{2}}.
\end{equation}
For a direct estimator $\mathcal{Z}^{0,1}$ this ratio takes the form
\begin{equation}
\frac{\langle(\mathcal{Z}^{0,1})^{2}\rangle^{(0)}}{(\langle\mathcal{Z}%
^{0,1}\rangle^{(0)})^{2}}=\frac{Q(0)\int(\mathcal{Z}^{0,1})^{2}\rho
^{(0)}(\mathbf{r})d\mathbf{r}}{[Q(1)]^{2}}. \label{eq:relative_RMSE}%
\end{equation}

For $\mathcal{Z}_{\mathrm{th}}^{0,1},$ we have%
\begin{widetext}
\begin{align}
\begin{split}
\int(\mathcal{Z}_{\mathrm{th}}^{0,1})^{2}\rho^{(0)}(\mathbf{r})d\mathbf{r}= &
C\int d\mathbf{r\,}\left[  \prod_{i=1}^{N}\frac{m_{i}(1)}{m_{i}(0)}\right]
^{P}\\
&  \times\exp\left\{  \frac{P}{2\beta\hbar^{2}}\sum_{i=1}^{N}[m_{i}%
(0)-2m_{i}(1)]\sum_{s=1}^{P}|\mathrm{r}_{i}^{(s)}-\mathrm{r}_{i}^{(s-1)}%
|^{2}-\frac{\beta}{P}\sum_{s=1}^{P}V(\mathbf{r}^{(s)})\right\}
\end{split}\label{eq:Z_th_2_int}
\\
\begin{split}
= &  \left(  \prod_{i=1}^{N}\left\{  1+\frac{[m_{i}(1)-m_{i}(0)]^{2}}%
{m_{i}(0)[2m_{i}(1)-m_{i}(0)]}\right\}  \right)  ^{P/2}\left(  \frac{P}%
{2\pi\beta\hbar^{2}}\right)  ^{NP/2}\left\{  \prod_{i=1}^{N}[2m_{i}%
(1)-m_{i}(0)]\right\}  ^{P/2}\\
&  \times\int\exp\left\{  -\frac{P}{2\beta\hbar^{2}}\sum_{i=1}^{N}%
[2m_{i}(1)-m_{i}(0)]\sum_{s=1}^{P}|\mathrm{r}_{i}^{(s)}-\mathrm{r}_{i}%
^{(s-1)}|^{2}-\frac{\beta}{P}\sum_{s=1}^{P}V(\mathbf{r}^{(s)})\right\}
d\mathbf{r}.
\end{split}
\end{align}
If $m_{i}(0)>2m_{i}(1)$ for some $i$, then the first term in the argument of the exponential becomes positive and for large enough values of $P$ the integral over $\mathbf{r}$ starts to diverge; this can be easily seen after rewriting it in terms of mass-scaled normal modes (see Appendix~A of Ref.~\onlinecite{Karandashev_Vanicek:2017a}). If $m_{i}(0)<2m_{i}(1)$, then interpreting the integral over $\mathbf{r}$ as the path integral representation of the partition function~(\ref{eq:Qr_PI}) for a harmonic system with rescaled masses reveals the limiting behavior
\begin{equation}%
\begin{split}
\lim_{P\rightarrow\infty}\int(\mathcal{Z}_{\mathrm{th}}^{0,1})^{2}\rho
^{(0)}(\mathbf{r})d\mathbf{r} &  =\lim_{P\rightarrow\infty}\prod_{i=1}%
^{N}\frac{\left\{  1+[m_{i}(1)-m_{i}(0)]^{2}/m(0)/[2m_{i}(1)-m_{i}
(0)]\right\}  ^{P/2}}{2\sinh\left( \beta\hbar\omega_{i}\sqrt
{m(0)/[2m_{i}(1)-m_{i}(0)]}/2\right)  }\\
&  =+\infty,
\end{split}
\end{equation}
implying the \emph{divergence of the statistical error of the thermodynamic direct estimator}, consistent with the similar divergence of the ``thermodynamic free energy perturbation'' estimator of Ref.~\onlinecite{Ceriotti_Markland:2013}. For $\mathcal{Z}%
_{\mathrm{sc}}^{0,1}$, we have
\begin{equation}%
\begin{split}
\int(\mathcal{Z}_{\mathrm{sc}}^{0,1})^{2}\rho(\mathbf{r})d\mathbf{r}=
C\left[  \prod_{i=1}^{N}\frac{m_{i}(1)}{m_{i}(0)}\right]  \int d\mathbf{r}\exp\left\{  -\frac{P}{2\beta\hbar^{2}}\sum_{i=1}^{N}m_{i}%
(0)\sum_{s=1}^{P}|\mathrm{r}_{i}^{(s)}-\mathrm{r}_{i}^{(s-1)}|^{2}-\frac
{\beta}{P}\sum_{s=1}^{P}\left[  2V(\mathbf{r}_{0,1}^{(s)})-V(\mathbf{r}%
^{(s)})\right]  \right\}  .
\end{split}
\end{equation}
To simplify this expression, let us rewrite it in terms of mass-scaled normal modes $\mathbf{u}$ (see Appendix~A of
Ref.~\onlinecite{Karandashev_Vanicek:2017a}). We will only consider the case of an even $P$, with the case of odd $P$ being completely analogous. For a harmonic system, we have
\begin{equation}
\begin{split}
\frac{1}{P}\sum_{s=1}^{P}V(\mathbf{r}^{(s)}) &  =\sum_{i=1}^{N}\frac
{m_{i}(0)\omega_{i}^{2}}{2P}\sum_{s=1}^{P}|\mathrm{r}_{i}^{(s)}|^{2}\\
&  =\sum_{i=1}^{N}\omega_{i}^{2}\left[  \frac{m_{i}(0)|\mathrm{r}_{i}%
^{(C)}|^{2}}{2}+\frac{|a_{i}^{(P/2)}|^{2}}{2}+\sum_{k=1}^{P/2-1}(|a_{i}%
^{(k)}|^{2}+|b_{i}^{(k)}|^{2})\right] ,
\end{split}
\end{equation}
and recalling the expression~(\ref{eq:rescaled_r_defined}) for $\mathbf{r}_{0,1}^{(s)}$  in terms of mass-scaled normal modes yields
\begin{equation}
\frac{1}{P}\sum_{s=1}^{P}V(\mathbf{r}_{0,1}^{(s)})=\sum_{i=1}^{N}\omega
_{i}^{2}\frac{m_{i}(0)}{m_{i}(1)}\left[  \frac{m_{i}(1)|\mathrm{r}_{i}%
^{(C)}|^{2}}{2}+\frac{|a_{i}^{(P/2)}|^{2}}{2}+\sum_{k=1}^{P/2-1}(|a_{i}%
^{(k)}|^{2}+|b_{i}^{(k)}|^{2})\right]  .
\end{equation}
Combining this with the expression for the effective potential
$\tilde{\Phi}(\mathbf{u})$ (see Appendix~A of
Ref.~\onlinecite{Karandashev_Vanicek:2017a}) leads to
\begin{equation}%
\begin{split}
\int(\mathcal{Z}_{\mathrm{sc}}^{0,1})^{2}\rho(\mathbf{u})d\mathbf{u}= &
\left[  \prod_{i=1}^{N}\frac{m_{i}(1)}{m_{i}(0)}\right]  \tilde{C}\int
d\mathbf{r}^{(C)}\exp\left[  -\beta\sum_{i=1}^{N}\frac{m_{i}(0)\omega_{i}%
^{2}|\mathrm{r}_{i}^{(C)}|^{2}}{2}\right]  \\
&  \times\int d\mathbf{a}^{(P/2)}\exp\left\{  -\frac{2P^{2}|\mathbf{a}%
^{(P/2)}|^{2}}{\beta\hbar^{2}}-\beta\sum_{i=1}^{N}\frac{\omega_{i}%
^{2}|\mathrm{a}_{i}^{(P/2)}|^{2}}{2}\left[  2\frac{m_{i}(0)}{m_{i}%
(1)}-1\right]  \right\}  \\
&  \times\prod_{k=1}^{P/2-1}\int d\mathbf{a}^{(k)}d\mathbf{b}^{(k)}%
\exp\left\{  -\frac{2P^{2}}{\beta\hbar^{2}}\left(  1-\cos\frac{2\pi k}%
{P}\right)  (|\mathbf{a}^{(k)}|^{2}+|\mathbf{b}^{(k)}|^{2})\right.  \\
&  \left.  -\beta\sum_{i=1}^{N}\frac{\omega_{i}^{2}(|\mathrm{a}_{i}^{(k)}%
|^{2}+|\mathrm{b}_{i}^{(k)}|^{2})}{2}\left[  2\frac{m_{i}(0)}{m_{i}%
(1)}-1\right]  \right\}  .
\end{split}
\label{eq:Z_v_second_moment}%
\end{equation}
If $m_{i}(1)>2m_{i}(0)$ for some $i$, then for large enough $\omega_{i}$ the
expression will diverge, implying the \emph{divergence of the statistical
error of the mass-scaled direct estimator}. Otherwise rescaling each
$\mathrm{r}_{i}^{(C)}$ by a factor of $\sqrt{2m_{i}(0)/m_{i}(1)-1}$ and
comparing the resulting expression to $Q_{P}$ rewritten in terms of
$\mathbf{u}$ yields
\begin{equation}
\lim_{P\rightarrow\infty}\int(\mathcal{Z}_{\mathrm{sc}}^{0,1})^{2}%
\rho(\mathbf{u})d\mathbf{u}=\prod_{i=1}^{N}\left\{  \frac{m_{i}(1)}{m_{i}%
(0)}\frac{\sqrt{2m_{i}(0)/m_{i}(1)-1}}{2\sinh(\beta\hbar\omega_{i}\sqrt
{2m_{i}(0)/m_{i}(1)-1}/2)}\right\}
\end{equation}
\end{widetext}
We now introduce the function
\begin{equation}
f_{i}(x)=\frac{\sqrt{x}}{2\sinh(\beta\hbar\sqrt{k_{i}x}/2)}%
,\label{eq:function_f}%
\end{equation}
where $k_{i}$ is the force constant for the $i$th degree of freedom, in order to rewrite expression~(\ref{eq:relative_RMSE}) in the limit
$P\rightarrow\infty$ as
\begin{equation}
\frac{\langle(\mathcal{Z}_{\mathrm{sc}}^{0,1})^{2}\rangle^{(0)}}%
{(\langle\mathcal{Z}_{\mathrm{sc}}^{0,1}\rangle^{(0)})^{2}}=\prod_{i=1}%
^{N}\frac{f_{i}[1/m_{i}(0)]f_{i}[2/m_{i}(1)-1/m_{i}(0)]}{\{f_{i}%
[1/m_{i}(1)]\}^{2}}.\label{eq:RMSE_relative_from_f}%
\end{equation}
Since $f_{i}(x)$ is a monotonously decreasing function of $x$ for each
$i=1,\ldots,N$, it is straightforward to show that if $m_{i}(0)<m_{i}(1)$ for
all $i=1,\ldots,N$, then the identity (\ref{eq:RMSE_relative_from_f}) implies
the inequality%
\begin{equation}
\frac{\langle(\mathcal{Z}_{\mathrm{sc}}^{1,0})^{2}\rangle^{(1)}}%
{(\langle\mathcal{Z}_{\mathrm{sc}}^{1,0}\rangle^{(1)})^{2}}<\frac
{\langle(\mathcal{Z}_{\mathrm{sc}}^{0,1})^{2}\rangle^{(0)}}{(\langle
\mathcal{Z}_{\mathrm{sc}}^{0,1}\rangle^{(0)})^{2}}%
.\label{eq:large_small_isotope_inequality}%
\end{equation}
This final inequality proves, for harmonic systems, an observation from Ref.~\onlinecite{Cheng_Ceriotti:2014}: \emph{If one evaluates an isotope
effect in a harmonic system using direct estimators for the isotope effect,
then running a path integral simulation at the larger value of mass leads to
smaller statistical errors than running the simulation at the lower value of
mass.} 

\section{\label{app:ODE_conv}Sufficient conditions for the convergence of direct estimators for
isotope substitution}

In this appendix we discuss several sufficient conditions that guarantee that
direct estimators exhibit a finite root mean square error. For $\mathcal{Z}%
_{\mathrm{th}}^{0,1}$ this is obviously the case if $2m_{i}(1)\geq m_{i}(0)$
for all $i=1,\ldots,N$, as in this case the path integral appearing in
Eq.~(\ref{eq:Z_th_2_int}) will have an upper bound of
\begin{equation}
\int(\mathcal{Z}_{\mathrm{th}}^{0,1})^{2}\rho^{(0)}(\mathbf{r})d\mathbf{r}\leq
C\left[\prod_{i=1}^{N}\frac{m_{i}(1)}{m_{i}(0)}\int e^{-\beta V(\mathbf{r}%
^{(1)})/P}d\mathbf{r}^{(1)}\right]^{P}.
\end{equation}
This upper bound goes to infinity as $ P\rightarrow\infty$, as should be expected from $ \mathcal{Z}_{\mathrm{th}}^{0,1}$ exhibiting an infinitely large statistical error in this limit. For $\mathcal{Z}_{\mathrm{sc}}^{0,1}$, we start by noting that the root mean square error of a bound observable is always finite, which is obviously the case if the potential function $V$ itself is bound. $\mathcal{Z}_{\mathrm{sc}}^{0,1}$ is also bound if $V$ is convex and $m_{i}(0)>m_{i}(1)$ for all $i=1,\ldots,N$. Indeed, for a given $ s=1,\ldots,P$
\begin{equation}
 V(\mathbf{r}^{(s)})-V(\mathbf{r}^{(s)}_{0,1})\leq \nabla V(\mathbf{r}^{(s)})\cdot(\mathbf{r}^{(s)}-\mathbf{r}^{(s)}_{0,1}),
 \label{eq:convexity_def}
\end{equation}
which follows from $ V(\mathbf{r}^{(s)})$ being a convex function of one variable when
change along the direction $ \mathbf{r}^{(s)}-\mathbf{r}^{(s)}_{0,1}$ is considered. From the definition~(\ref{eq:rescaled_r_defined}) of $ \mathbf{r}^{(s)}_{0,1}$ we obtain a positive definite
matrix $ \mathbf{A}$ such that
\begin{equation}
 \mathbf{r}^{(s)}-\mathbf{r}^{(s)}_{0,1}=\mathbf{A}\cdot(\mathbf{r}^{(s)}-\mathbf{r}^{(C)}).
\end{equation}
We also introduce 
\begin{equation}
 \tilde{\mathbf{r}}^{(s)}(\tau)=\tau\mathbf{r}^{(s)}+(1-\tau)\mathbf{r}^{(C)}
\end{equation}
and write
\begin{widetext}
\begin{equation}
\begin{split}
 \nabla V(\mathbf{r}^{(s)})\cdot\mathbf{A}\cdot(\mathbf{r}^{(s)}-\mathbf{r}^{(C)})=&
 \nabla V(\mathbf{r}^{(C)})\cdot\mathbf{A}\cdot(\mathbf{r}^{(s)}-\mathbf{r}^{(C)})+\int_{0}^{1}d\tau
 (\mathbf{r}^{(s)}-\mathbf{r}^{(C)})\cdot\left(\mathbf{A}\left.\frac{\partial^{2}
 V(\mathbf{r}^{(s)})}{\partial(\mathbf{r}^{(s)})^{2}}\right|_{\mathbf{r}^{(s)}=\tilde{\mathbf{r}}^{(s)}(\tau)}\right)
 \cdot(\mathbf{r}^{(s)}-\mathbf{r}^{(C)}).
\end{split}
\label{eq:tau_integral}
\end{equation}
\end{widetext}
Since $ V$ is a convex function, $ \partial^{2}V(\mathbf{r}^{(s)})/\partial(\mathbf{r}^{(s)})^{2}$ is
a positive semi-definite matrix, and therefore $ \mathbf{A}\partial^{2}V(\mathbf{r}^{(s)})/\partial(\mathbf{r}^{(s)})^{2}$
has non-negative eigenvalues. Thus, for each $ \tau$
\begin{equation}
 (\mathbf{r}^{(s)}-\mathbf{r}^{(C)})\cdot\left(\mathbf{A}\left.\frac{\partial^{2}
 V(\mathbf{r}^{(s)})}{\partial(\mathbf{r}^{(s)})^{2}}\right|_{\mathbf{r}^{(s)}=\tilde{\mathbf{r}}^{(s)}(\tau)}\right)
 \cdot(\mathbf{r}^{(s)}-\mathbf{r}^{(C)})\geq0.
 \label{eq:matrix_diag_element_ineq}
\end{equation}
Combining Eqs.~(\ref{eq:convexity_def}), (\ref{eq:tau_integral}), and (\ref{eq:matrix_diag_element_ineq}),
then summing the result over $ s$ leads us to
\begin{equation}
 \sum_{s=1}^{P}[V(\mathbf{r}^{(s)})-V(\mathbf{r}^{(s)}_{0,1})]\leq
 \nabla V(\mathbf{r}^{(C)})\cdot\mathbf{A}\cdot\sum_{s=1}^{P}(\mathbf{r}^{(s)}-\mathbf{r}^{(C)})=0,
\end{equation}
proving that in this case $ \mathcal{Z}_{\mathrm{sc}}^{0,1}$ is bound in a way that does not depend on $ P$
\begin{equation}
 \mathcal{Z}_{\mathrm{sc}}^{0,1}\leq \left[  \prod_{i=1}^{N}\frac{m_{i}(1)}%
{m_{i}(0)}\right]  ^{D/2}.
\end{equation}

\section{\label{app:DE_err}Optimal choice of mass discretization and reference masses for the stepwise application of direct estimators}

In this appendix we discuss choices of $\lambda_{j}$ and $\overline{\lambda}_{j}$ that minimize the root mean square error of the direct estimator expression~(\ref{eq:DE_stepwise}). In other words, we discuss the optional choice of the partition of the interval $[0,1]$ into subintervals $[\lambda_{j-1},\lambda_{j}]$ and the optimal choice of the reference masses $(\overline{\lambda}_{j})$ in each of the subintervals $[\lambda_{j-1},\lambda_{j}]$.

The first part of the problem is the choice
of $\overline{\lambda}$ that minimizes the statistical error of the ratio $\langle\mathcal{Z}_{\mathrm{sc}}^{\overline{\lambda},1}\rangle^{(\overline
{\lambda})}/\langle\mathcal{Z}_{\mathrm{sc}}^{\overline{\lambda},0}\rangle^{(\overline{\lambda})}$. Estimating the root mean square error of this ratio analytically is problematic as the averages are obtained from the same Monte Carlo trajectory, and hence are correlated. Even if one could neglect
the correlation between the two averages, the resulting estimate for an optimal $\overline{\lambda}$ value would be too complicated to be useful. Yet,
we can suggest two rules of thumb based on an approximate behavior in the deep quantum regime because the most quantum degrees of freedom typically
contribute the most to the statistical error. [Equation~(\ref{eq:RMSE_relative_from_f}) can be shown to imply this.] For
simplicity let us consider a one-dimensional harmonic oscillator with force
constant $k$, mass changed from $m(0)$ to $m(1)>m(0)$, and $m(\lambda)$ given
by the inverse square root interpolation formula~(\ref{eq:m_root_interpol}).
Assuming the averages of $\mathcal{Z}_{\mathrm{sc}}^{\overline{\lambda},1}$
and $\mathcal{Z}_{\mathrm{sc}}^{\overline{\lambda},0}$ to be approximately
independent and to have the limiting behavior of Eq.~(\ref{eq:relative_RMSE}) at
low temperatures allows us to write
\begin{widetext}
\begin{equation}
\begin{split}
\left[  \frac{\Delta(\langle\mathcal{Z}_{\mathrm{sc}}^{\overline{\lambda}%
,1}\rangle^{(\overline{\lambda})}/\langle\mathcal{Z}_{\mathrm{sc}}%
^{\overline{\lambda},0}\rangle^{(\overline{\lambda})})}{\langle\mathcal{Z}%
_{\mathrm{sc}}^{\overline{\lambda},1}\rangle^{(\overline{\lambda})}%
/\langle\mathcal{Z}_{\mathrm{sc}}^{\overline{\lambda},0}\rangle^{(\overline
{\lambda})}}\right]  ^{2}   \approx\,\,\,\,\,\,\,&\left[  \frac{\langle(\mathcal{Z}%
_{\mathrm{sc}}^{\overline{\lambda},1})^{2}\rangle}{\langle\mathcal{Z}%
_{\mathrm{sc}}^{\overline{\lambda},1}\rangle^{2}}+\frac{\langle(\mathcal{Z}%
_{\mathrm{sc}}^{\overline{\lambda},0})^{2}\rangle}{\langle\mathcal{Z}%
_{\mathrm{sc}}^{\overline{\lambda},0}\rangle^{2}}-2\right]  \\
  \stackrel{\mathclap{\normalfont\mbox{\footnotesize $ \beta\rightarrow\infty$}}}{\sim}\,\,\,\,\,\,\,&\frac{\sqrt{m(1)[2m(\overline{\lambda})-m(1)]}%
}{m(\overline{\lambda})}\exp\left\{  \beta\hbar\sqrt{k}\left[  \frac{1}{\sqrt{m(1)}}-\frac
{1}{2\sqrt{m(\overline{\lambda})}}-\frac{1}{2}\sqrt{\frac{2}{m(1)}-\frac{1}{m(\overline{\lambda})}%
}\right]  \right\}  \\
&  +\frac{\sqrt{m(0)[2m(\overline{\lambda})-m(0)]}}{m(\overline{\lambda})}\exp\left\{  \beta\hbar\sqrt{k}\left[  \frac{1}{\sqrt{m(0)}}-\frac
{1}{2\sqrt{m(\overline{\lambda})}}-\frac{1}{2}\sqrt{\frac{2}{m(0)}-\frac{1}{m(\overline{\lambda})}%
}\right]  \right\}.
\end{split}
\label{eq:relative_RMSE_low_temperature_limit}%
\end{equation}
\end{widetext}
We also assume that $ \mathcal{Z}_{\mathrm{sc}}^{\overline{\lambda},1}$ does not exhibit a divergent behavior [as described in Appendix~\ref{app:ODE_err}], since it could only appear in this problem for very exotic mass differences [$ m(1)>2m(0)$] and a too sparse discretization into $ \lambda$ subintervals. In the $ \beta\rightarrow\infty$ limit, differentiating either term in~(\ref{eq:relative_RMSE_low_temperature_limit}) with respect to $ \overline{\lambda}$ and dividing the result by $ \beta\mathrm{d}m(\overline{\lambda})/\mathrm{d}\overline{\lambda}$ leads to an expression whose magnitude in the $ \beta\rightarrow\infty$ limit will be mainly determined by the argument of the exponential. The minimum corresponds to derivatives of the two terms having the same magnitude to cancel out, which implies the arguments of the two exponentials being approximately equal. Straightforward algebra indicates the argument of the first exponent is larger than in the second one at $\overline{\lambda}=1/2$, while the inequality is reversed at $\overline{\lambda}=1 $, implying that the minimum lies in the interval $[1/2,1]$, which is consistent
with the optimal $\bar{\lambda}$ value found in test calculations we have performed with several harmonic systems. We have not observed a significant difference between statistical errors obtained for $\overline{\lambda}=1/2$ or $\overline{\lambda}=1$ in a wide range of situations; in the
limit of large $J$, however, picking $\overline{\lambda}=1/2$ becomes
numerically identical to performing thermodynamic integration with the midpoint rule, while choosing $\overline{\lambda}=1$ would correspond to the right hand rule, which should lead to worse performance. $\overline{\lambda}=1/2$ was therefore the choice used in this work.

\begin{figure}
\centering\includegraphics[width=0.375\textwidth, trim={0 0 8.4cm 0}]{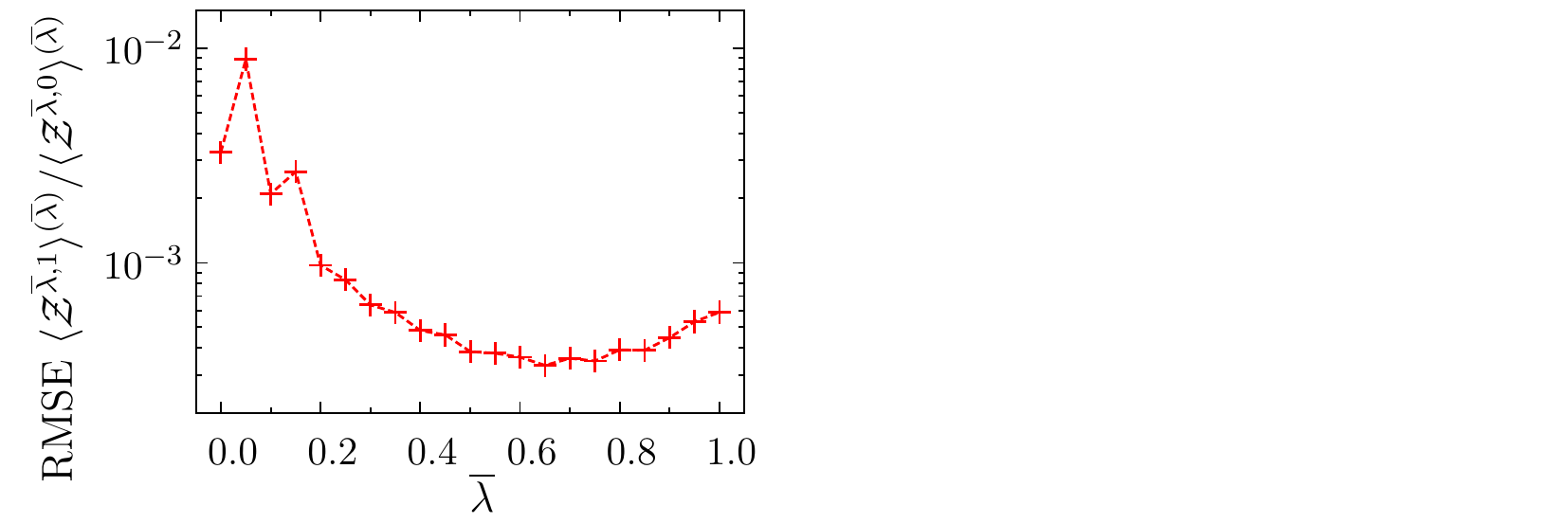}
\caption{\label{fig:sho_over_l_stat_err}Relative root mean square error (RMSE)
of $  \langle\mathcal{Z}^{\overline{\lambda},1}\rangle^{(\overline{\lambda})}
/\langle\mathcal{Z}^{\overline{\lambda},0}\rangle^{(\overline{\lambda})}$ as a function of $ \overline{\lambda}$
for the harmonic system used in Ref.~\onlinecite{Karandashev_Vanicek:2017a}
with $ \beta\hbar\omega_{0}=32$.}
\end{figure}

To illustrate our estimate for the optimal $\overline{\lambda}$, in
Fig.~\ref{fig:sho_over_l_stat_err} we plot root mean square error of
$\langle\mathcal{Z}^{\overline{\lambda},1}\rangle^{(\overline{\lambda}%
)}/\langle\mathcal{Z}^{\overline{\lambda},0}\rangle^{(\overline{\lambda})}$ as
a function of $\overline{\lambda}$ for the same harmonic system as in
Subsection~\ref{subsec:SDE_harmonic} with $\beta\hbar\omega_{0}=32$.
Evidently, the optimal $\overline{\lambda}$ is in the interval $[1/2,1]$ and
is quite close to $1/2$.

\begin{figure}
\centering\includegraphics[width=0.375\textwidth, trim={0 0 8.4cm 0}]{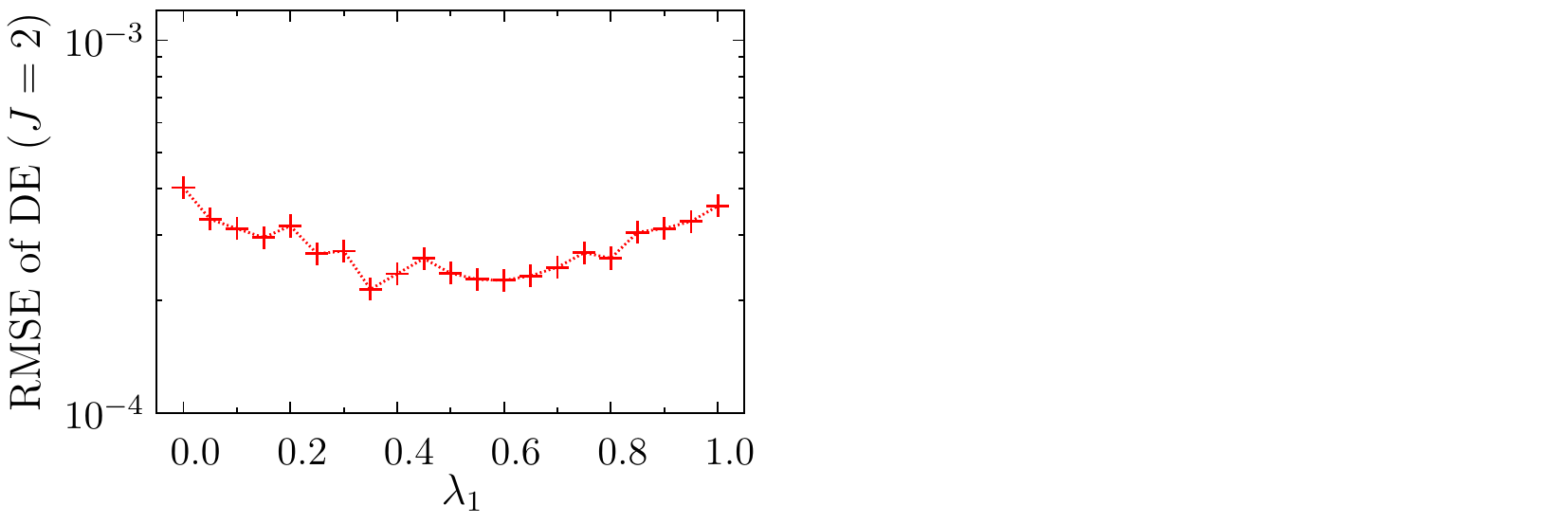}
\caption{\label{fig:sho_middle_stat_err}Relative root mean square error (RMSE)
of the isotope effect evaluated with direct estimators~[DE, Eq.~(\ref{eq:DE_stepwise})] with $ J=2$
as a function of $ \lambda_{1}$
for the harmonic system used in Ref.~\onlinecite{Karandashev_Vanicek:2017a}
with $ \beta\hbar\omega_{0}=32$.}
\end{figure}

We will now discuss the best way to partition a large isotope effect into smaller isotope effects, each of which can be evaluated with the original direct estimators. Assuming that the midpoint was used for an evaluation of an intermediate isotope effect $Q(\lambda^{\prime\prime})/Q(\lambda^{\prime})$ one can obtain the following approximate upper bound for the relative statistical error in the low temperature limit
\begin{widetext}
\begin{equation}
\left[  \frac{\Delta(\langle\mathcal{Z}_{\mathrm{sc}}^{\overline{\lambda}%
,1}\rangle^{(\overline{\lambda})}/\langle\mathcal{Z}_{\mathrm{sc}}%
^{\overline{\lambda},0}\rangle^{(\overline{\lambda})})}{\langle\mathcal{Z}%
_{\mathrm{sc}}^{\overline{\lambda},1}\rangle^{(\overline{\lambda})}%
/\langle\mathcal{Z}_{\mathrm{sc}}^{\overline{\lambda},0}\rangle^{(\overline
{\lambda})}}\right]  ^{2}  \lesssim_{\beta\rightarrow\infty}\exp\left\{
\frac{\beta\hbar\sqrt{k}}{2}\left[  \frac{1}{\sqrt{m(0)}}-\frac{1}{\sqrt
{m(1)}}\right]  \right\}  \approx\frac{Q(1)}{Q(0)}.
\label{eq:relative_RMSE_low_temperature_upper_bound}%
\end{equation}
\end{widetext}
In this approximation the problem of minimizing statistical error of a
\textquotedblleft stepwise\textquotedblright\ direct estimator isotope effect
calculation is equivalent to minimizing a sum of positive terms whose product is
fixed. Since the solution of the latter problem requires all terms of the sum being
equal, the resulting rule of thumb is to pick the intermediate isotope effects
approximately equal in magnitude. In the low temperature limit, and with mass interpolation~(\ref{eq:m_root_interpol}), this is
equivalent to having the intermediate isotope effects correspond to $\lambda$
subintervals of equal size. To check this estimate we ran some test
calculations for the same system as in Subsection~\ref{subsec:SDE_harmonic}
with $\beta\hbar\omega_{0}=32$ using direct estimators and $J=2$ with
different values of $\lambda_{1}$. The resulting root mean square errors are
plotted in Fig.~\ref{fig:sho_middle_stat_err}; for this rather quantum
harmonic oscillator root mean square error is indeed minimized if $\lambda
_{1}$ is close to $1/2$.

\section{\label{app:SDE_details}Details of combining direct estimators with
stochastic change of mass}

The procedure for changing masses between discrete values $\overline{\lambda
}_{j}$ of $\lambda$ for SDE is mostly the same as the one described for STI in
Subsection~II C of Ref.~\onlinecite{Karandashev_Vanicek:2017a}, but for two
important differences. First, the umbrella potential $U_{b}(\lambda)$ is now
updated in such a way that it satisfies for all $j=1,\ldots J-1$ the
condition
\begin{equation}
\exp\{\beta\lbrack U_{b}(\overline{\lambda}_{j})-U_{b}(\overline{\lambda
}_{j+1})]\}=\frac{\langle\mathcal{Z}_{\mathrm{sc}}^{\overline{\lambda}%
_{j+1},\lambda_{j}}\rangle^{(\overline{\lambda}_{j+1})}}{\langle
\mathcal{Z}_{\mathrm{sc}}^{\overline{\lambda}_{j},\lambda_{j}}\rangle
^{(\overline{\lambda}_{j})}} \label{eq:SDE_U_b_condition}%
\end{equation}
in order to minimize the statistical error of the calculated isotope effect.
Second, since the evaluation of the acceptance criterion for the simple
$\lambda$-move appearing in Eq.~(23) of
Ref.~\onlinecite{Karandashev_Vanicek:2017a} is computationally inexpensive and
we only need to consider a finite (and typically small) number of $\lambda
$-values, it is possible to include the acceptance probability into the trial
distribution, leading to the following procedure:

\textbf{Simple $\lambda$-move:}

\begin{enumerate}
\setlength{\itemindent}{0.05\textwidth}

\item For each $j=1,\ldots,J$ calculate:
\begin{align}
\begin{split}
p_{j} = & \left[  \prod_{i=1}^{N}m_{i}(\overline{\lambda}_{j})\right]\\
&\times\exp\left[  -\frac{P}{2\beta\hbar^{2}}\sum_{i=1}^{N}m_{i}(\overline{\lambda
}_{j})\sum_{s=1}^{P}|\mathrm{r}_{i}^{(s)}-\mathrm{r}_{i}^{(s-1)}|^{2} \right.\\
&\left.\vphantom{-\frac{P}{2\beta\hbar^{2}}\sum_{i=1}^{N}m_{i}(\overline{\lambda
}_{j})\sum_{s=1}^{P}|\mathrm{r}_{i}^{(s)}-\mathrm{r}_{i}^{(s-1)}|^{2}}-\beta
U_{b}(\overline{\lambda}_{j})\right]
 \label{eq:simple_l_move_p_def}
\end{split}\\
\tilde{p}_{j} =&\frac{\sum_{j^{\prime}=1}^{j}p_{j^{\prime}}}{\sum
_{j^{\prime}=1}^{J}p_{j^{\prime}}}%
\end{align}

\item Choose a random number $\Delta$ distributed uniformly over $[0,1]$.

\item Choose $j^{\prime}$ which is the smallest integer satisfying
\begin{equation}
\Delta<\tilde{p}_{j^{\prime}}.
\end{equation}

\item Set $j=j^{\prime}$, $\lambda=\overline{\lambda}_{j^{\prime}}$.
\end{enumerate}

It was mentioned in Subsection~II C of
Ref.~\onlinecite{Karandashev_Vanicek:2017a} that {simple $\lambda$-moves used
for stochastic thermodynamic integration cannot lead} to large changes of
$\lambda$ as the\textbf{ }corresponding acceptance ratio [Eq.~(23) of
Ref.~\onlinecite{Karandashev_Vanicek:2017a}, which is basically the ratio of
two different $p_{j}$ from Eq.~(\ref{eq:simple_l_move_p_def})] exhibits a maximum which becomes sharper with larger values of $P$. We note that the SDE variant of the
procedure outlined above cannot address this issue, as the smallest $\lambda$
step one can make is limited by $J$, and therefore becomes less efficient at
lower temperatures.

Finally, as for the mass-scaled $\lambda$-move, the only difference from the
procedure for STI in Ref.~\onlinecite{Karandashev_Vanicek:2017a} is that, in
the case of SDE, the trial $\lambda$ values [Eq.~(24)
Ref.~\onlinecite{Karandashev_Vanicek:2017a}] are picked not from the entire
$[0,1]$ interval, but only from among the reference $\overline{\lambda}_{j}$ values.

\section{\label{app:discr_error}Estimate of the relative path integral
discretization error}

To estimate the discretization error of the path integral representation
$Q_{P}$ of the partition function we start from the identity,
\begin{equation}
Q_{P}=Q+\mathcal{O}\left(  \frac{1}{P^{n}}\right)  =Q+\frac{c}{P^{n}}+o\left(
\frac{1}{P^{n}}\right)  , \label{eq:absolute_discr_error_expression}%
\end{equation}
where $c$ is independent of $P$ and the integer $n$ depends on the
factorization used to derive $Q_{P}$ [in particular, $n=2$ for the Lie-Trotter
and $n=4$ for Takahashi-Imada \cite{Takahashi_Imada:1984} and Suzuki-Chin
(SC)\cite{Suzuki:1995,Chin:1997} factorizations]. As usual, the little-o
symbol is defined by the relation $f(x)=o[g(x)]$ if $g(x)\neq0$ in some
neighborhood of $x=0$ and $\lim_{x\rightarrow0}f(x)/g(x)=0$. We proceed to
rewrite the relative discretization error of $Q_{P}$ as
\begin{widetext}
\begin{equation}%
\begin{split}
\frac{Q_{P}-Q}{Q}  &  =\frac{c/P^{n}+o(P^{-n})}{Q}=\frac{1}{1-2^{-n}}%
\frac{Q+c/P^{n}-Q-c/(2P)^{n}+o(P^{-n})}{Q_{P}+\mathcal{O}(P^{-n})}\\
&  =\frac{1}{1-2^{-n}}\frac{Q_{P}-Q_{2P}+o(P^{-n})}{Q_{P}+\mathcal{O}(P^{-n}%
)}=\frac{1}{1-2^{-n}}\frac{Q_{P}-Q_{2P}}{Q_{P}}+o(P^{-n}).
\end{split}
\label{eq:discr_error_est}%
\end{equation}
It follows that we can estimate the discretization error of $Q_{P}$ if we can
estimate the ratio $Q_{2P}/Q_{P}$. We shall therefore derive a direct
estimator for $Q_{2P}/Q_{P}$; for simplicity, we will do this explicitly only
for the special case of Lie-Trotter splitting ($n=2$). The derivation, which
resembles that presented in Ref.~\onlinecite{Azuri_Major:2011} for the direct
estimator of $Q_{P}/Q_{1}$, starts by expressing $Q_{2P}$ as
\begin{equation}
\begin{split}
Q_{2P}=  &  \left(  \frac{P}{\pi\beta\hbar^{2}}\right)  ^{DNP}\left(
\prod_{i=1}^{N}m_{i}\right)  ^{DP}\int d\mathbf{r}^{\prime}d\mathbf{r}\\
&  \times\exp\left\{  -\frac{P}{\beta\hbar^{2}}\sum_{s=0}^{P-1}\left(
||\mathbf{r}^{\prime(s)}-\mathbf{r}^{(s)}||_{+}^{2}+||\mathbf{r}^{\prime
(s)}-\mathbf{r}^{(s+1)}||_{+}^{2}\right)  -\frac{\beta}{2P}\sum_{s=1}%
^{P}\left[  V(\mathbf{r}^{\prime(s)})+V(\mathbf{r}^{(s)})\right]  \right\}  ,
\end{split}
\end{equation}
\end{widetext}
where both $\mathbf{r}$ and $\mathbf{r}^{\prime}$ are sets of $P$ vector
variables in the system's configuration space, $||\mathbf{r}^{(s)}||_{+}$
denotes the norm of a contravariant vector $\mathbf{r}^{(s)}$ in the system's
configuration space; it is evaluated as%
\begin{equation}
||\mathbf{r}^{(s)}||_{+}^{2}:=\sum_{i=1}^{N}m_{i}|\mathrm{r}_{i}^{(s)}|^{2},
\end{equation}
where $\mathrm{r}_{i}^{(s)}$ is the component of $\mathbf{r}^{(s)}$
corresponding to particle $i$. Being a norm induced by an inner product,
$||\cdots||_{+}$ satisfies the parallelogram law
\begin{widetext}
\begin{equation}
\begin{split}
||\mathbf{r}^{\prime(s)}-\mathbf{r}^{(s)}||_{+}^{2}+||\mathbf{r}^{\prime
(s)}-\mathbf{r}^{(s+1)}||_{+}^{2}=\frac{1}{2}(||2\mathbf{r}^{\prime
(s)}-\mathbf{r}^{(s+1)}-\mathbf{r}^{(s)}||_{+}^{2}
+||\mathbf{r}^{(s)}%
-\mathbf{r}^{(s+1)}||_{+}^{2}),
\end{split}
\end{equation}
which allows to rewrite $Q_{2P}$ as
\begin{equation}
Q_{2P}=C\int\mathcal{W}_{2}d\mathbf{r}\rho(\mathbf{r}),
\end{equation}
where $\mathcal{W}_{2}$, defined as
\begin{equation}
\begin{split}
\mathcal{W}_{2}=  &  \int\exp\left\{  \frac{\beta}{2P}\sum_{s=1}%
^{P}[V(\mathbf{r}^{(s)})-V(\mathbf{r}^{\prime(s)})]\right\}\prod_{s=1}^{P}\left[  \left(  \prod_{i=1}^{N}m_{i}\right)
^{D/2}\left(  \frac{2P}{\pi\beta\hbar^{2}}\right)  ^{DN/2}\exp\left(
-\frac{2P}{\beta\hbar^{2}}\left\vert \left\vert \mathbf{r}^{\prime(s)}%
-\frac{\mathbf{r}^{(s)}+\mathbf{r}^{(s+1)}}{2}\right\vert \right\vert _{+}%
^{2}\right)  d\mathbf{r}^{\prime(s)}\right]  ,
\end{split}
\label{eq:W_2}%
\end{equation}
\end{widetext}
is obviously the direct estimator for $Q_{2P}/Q_{P}$. $\mathcal{W}_{2}$ can be
evaluated by generating $\mathbf{r}^{\prime}$ with the Box-Muller method and
averaging the resulting exponential factor; the procedure is in a
way reminiscent of the last step of bisection path integral sampling
method.\cite{Ceperley:1995,Major_Gao:2005} Note that for large values of $P$
the Gaussians from which $\mathbf{r}^{\prime}$ are sampled are quite narrow,
making the sum $\sum_{s=1}^{P}[V(\mathbf{r}^{(s)})-V(\mathbf{r}^{\prime(s)})]$
approach $0$, which should in turn lead to a reasonably fast statistical
convergence of the estimator.

Incidentally, one can express the discretization error of $Q_{P}$ by
evaluating a direct estimator for $Q_{P/2}/Q_{P}$, which we therefore denote
$\mathcal{W}_{1/2}$. This estimator can be derived completely analogously,
with the result
\begin{equation}
\mathcal{W}_{1/2}=\exp\left\{  \frac{\beta}{P}\sum_{s=1}^{P/2}\left[
V(\mathbf{r}^{(2s-1)})-V(\mathbf{r}^{(2s)})\right]  \right\}  .
\label{eq:W_half}%
\end{equation}
Unlike $\mathcal{W}_{2}$, $\mathcal{W}_{1/2}$ does not require additional
evaluations of the potential energy; however, estimating discretization error
from $Q_{P}/Q_{P/2}$ would typically yield less accurate results than
estimating it from $Q_{2P}/Q_{P}$.

As for the discretization error of the isotope effect itself, one can
similarly obtain the estimate
\begin{equation}
\begin{split}
\frac{\mathrm{IE}_{P}}{\mathrm{IE}}-1&=\frac{1}{1-2^{-n}}\left(  1-\frac
{\mathrm{IE}_{2P}}{\mathrm{IE}_{P}}\right)  +o(P^{-n})\\
&=\frac{1}{1-2^{-n}%
}\left(  1-\frac{\left\langle \mathcal{W}_{2}\right\rangle ^{(1)}%
}{\left\langle \mathcal{W}_{2}\right\rangle ^{(0)}}\right)  +o(P^{-n}).
\label{eq:discr_error_IE_approx}%
\end{split}
\end{equation}
In this case it is necessary to calculate two averages, at $\lambda=0$ and $\lambda=1$, to obtain the discretization error estimate.

\begin{figure*}\centering\includegraphics[width=0.75\textwidth]{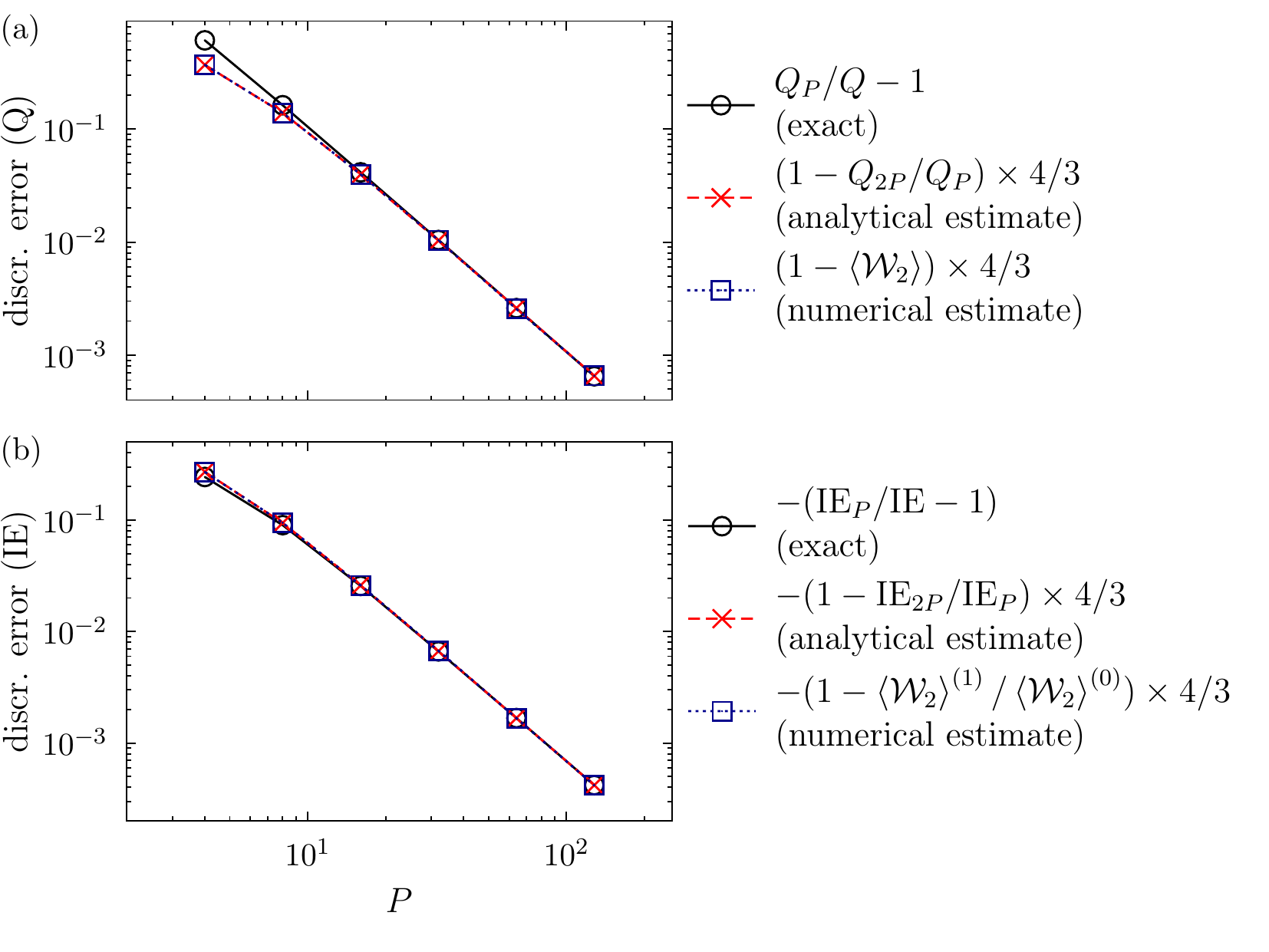}
\caption{\label{fig:SHO_disc_err}Comparison of exact analytical values of the path integral discretization error of (a) the partition function $Q$ and (b) isotope effect
with their estimates [Eqs.~(\ref{eq:discr_error_est}) and
(\ref{eq:discr_error_IE_approx})], which were evaluated either analytically
or numerically using the estimator~(\ref{eq:W_2}). The figure shows the dependence of the discretization error on the Trotter number $P$
in a one-dimensional harmonic oscillator with $\beta\hbar\omega=8$, and the isotope effect corresponds to the doubling of the mass.}
\end{figure*}

To make sure that our estimates are correct we ran test calculations for
one-dimensional harmonic oscillator with $\beta\hbar\omega=8$ at several $P$
values, with the isotope effect corresponding to the doubling of the mass. The
results, presented in Fig.~\ref{fig:SHO_disc_err}, show that our method for
estimating the discretization error becomes very accurate with increasing $P$,
and in fact could, in principle, be used to decrease the discretization error
of the calculation from $O(P^{-n})$ to $o(P^{-n})$ (by subtracting the error
estimate from the result). Unfortunately, a practical Monte Carlo calculation
also has a statistical error, which decreases only as an inverse square root
of the total number of Monte Carlo steps, while
Eq.~(\ref{eq:absolute_discr_error_expression}) implies that the discretization
error decreases approximately as $P^{-n}$. Since the total cost of the
calculation is approximately proportional to the product of the number of
Monte Carlo steps and $P$, it is clear that in a practical calculation the
statistical error will be much harder to decrease than the discretization
error. Therefore, in this manuscript, we opted to use the discretization error
estimate~(\ref{eq:discr_error_IE_approx}) only to make sure that the
discretization error of the isotope effect is comparable to statistical error.
However, it is also possible to subtract these discretization error estimates
from the calculated value of $\ln($IE$)$ in order to obtain a result that will
be closer to the quantum limit, but whose discretization error can no longer
be estimated. Thus the data presented in
Table~\ref{tab:lnIE_methane_values_and_disc_error} for the CD$_{4}/$CH$_{4}$
isotope effect allow two different ways to interpret the results; the choice
is left to the reader.

Finally, note that expressions analogous to Eqs. (\ref{eq:W_2}) and
(\ref{eq:W_half}) can also be derived for the fourth-order
Takahashi-Imada\cite{Takahashi_Imada:1984} and
Suzuki-Chin\cite{Suzuki:1995,Chin:1997} factorizations. One needs to be
careful, however, since both fourth-order factorization replace $V$ with an
effective potential $V_{\text{eff}}$ that, unlike $V$, depends on $P$. As for
the Suzuki-Chin factorization, the matter is further complicated by the fact
that the weight of this effective potential depends on the bead $s$ at which
it is evaluated.

\end{document}